\documentclass[%
 reprint,
superscriptaddress,
nofootinbib,
 amsmath,amssymb,
 aps,
 showkeys,
]{revtex4-1}
\usepackage{graphicx}% Include figure files
\usepackage{dcolumn}% Align table columns on decimal point
\usepackage{bm}% bold math
\usepackage{hyperref}
\usepackage[bottom]{footmisc}
\usepackage{footnote}
\usepackage{xcolor}
\usepackage[mathlines]{lineno}% Enable numbering of text and display math
\usepackage{fontenc}
\usepackage{graphicx}% Include figure files
\usepackage{dcolumn}% Align table columns on decimal point
\usepackage{bm}% bold math
\usepackage{booktabs}
\usepackage{array,multirow}
\usepackage{amsmath}
\usepackage{amsfonts}
\usepackage{amssymb}
\usepackage{mathrsfs}
\usepackage{enumerate}
\usepackage{fancyhdr}
\usepackage{xcolor}
\usepackage{graphicx}
\usepackage{listings}
\usepackage{float}
\usepackage{verbatim}
\usepackage{braket}
\usepackage{centernot}
\usepackage{setspace}
\usepackage{endnotes}
\usepackage[normalem]{ulem}
\usepackage{wrapfig}
\usepackage{multirow}
\usepackage{comment}
\usepackage{kantlipsum}
\allowdisplaybreaks
\usepackage{lipsum, babel}
\hypersetup{breaklinks = true, colorlinks = true, citecolor = blue, linkcolor = blue, urlcolor = blue}
\usepackage[mathlines]{lineno}% Enable numbering of text and display math
\newcommand{\be}{\begin{equation}}
\newcommand{\ee}{\end{equation}}
\newcommand{\bea}{\begin{eqnarray}}
\newcommand{\eea}{\end{eqnarray}}

\newcommand{\tp}{ \tilde{p}}

\begin{document}

\title{QCD evolution of entanglement entropy}% Force line breaks with \\
% 
% 
% please check your name and affiliations!!!!
% 
% 
\author{Martin~Hentschinski}
\email{martin.hentschinski@udlap.mx}
\affiliation{Departamento de Actuaria, F\'isica y Matem\'aticas, Universidad de las Am\'ericas Puebla, San Andres
Cholula, 72820 Puebla, Mexico}

\author{Dmitri E.~Kharzeev}
\email{dmitri.kharzeev@stonybrook.edu}
\affiliation{Center for Nuclear Theory, Department of Physics and Astronomy, Stony Brook University, New York 11794-3800, USA}
\affiliation{Department of Physics, Brookhaven National Laboratory, Upton, New York 11973, USA}

\author{Krzysztof~Kutak}
\email{krzysztof.kutak@ifj.edu.pl}
\affiliation{Institute of Nuclear Physics, Polish Academy of Sciences, ul. Radzikowskiego 152, 31-342,
Krak\'ow, Poland}

\author{Zhoudunming~Tu}
\email{zhoudunming@bnl.gov}
\affiliation{Department of Physics, Brookhaven National Laboratory, Upton, New York 11973, USA}

\date{\today}% It is always \today, today,
             %  but any date may be explicitly specified
\begin{abstract} 
Entanglement entropy has emerged as a novel tool for probing nonperturbative quantum chromodynamics (QCD) phenomena, such as color confinement in protons. While recent studies have demonstrated its significant capability in describing hadron production in deep inelastic scatterings, the QCD evolution of entanglement entropy remains unexplored. In this work, we investigate the differential rapidity-dependent entanglement entropy within the proton and its connection to final-state hadrons, aiming to elucidate its QCD evolution. Our analysis reveals a strong agreement between the rapidity dependence of von Neumann entropy, obtained from QCD evolution equations, and the corresponding experimental data on hadron entropy. These findings provide compelling evidence for the emergence of a maximally entangled state, offering new insights into the nonperturbative structure of protons.
\end{abstract}

%Use showkeys class option if keyword display desired
\maketitle
 
\section{Introduction}
{ Quantum entanglement, famously described by Albert Einstein as a "spooky action at a distance," is fundamental to our understanding of the microscopic world. The nontrivial and non-local correlations among the constituents of a quantum system can provide deep insights into the underlying properties of such systems. In recent years, there has been a growing interest in quantum entanglement within high-energy physics, particularly in understanding the structure of hadrons \cite{Kharzeev:2017qzs}, thermalization \cite{Alba:2017ekd} (see \cite{Berges:2020fwq} for a review), spin and helicity coupling of partons \cite{Bhattacharya:2024sno}, violation of Bell inequalities in hadronic final states \cite{Gong:2021bcp,Aoude:2022imd,Severi:2021cnj,White:2024nuc,CMS:2024pts,ATLAS:2023fsd,Banuls:2022iwk}, early universe QCD transition~\cite{Banerjee:2003fg}}, and confinement \cite{Klebanov:2007ws,Grieninger:2023ehb}.

To quantify entanglement of quantum systems one introduces entanglement measures \cite{Nielsen:2012yss,Horodecki:2009zz}. One of such measures is an entanglement entropy. The quantum system that is of interest to us is the proton and our main goal is to find entropy associated with proton and evolution of this entropy with energy. 
However, right at the beginning one faces the following issue: in quantum mechanics, a proton in isolation is a pure quantum state,
an eigenstate of the QCD Hamiltonian. However, a measurement performed on this state by an external probe often leads to a loss of information about the state. For example, in Deep Inelastic Scattering (DIS) at momentum transfer $t = -\,Q^2$, the virtual photon probes only a small fraction of the proton's wave function with a transverse area $\sim 1/Q^2$. Moreover, in any high-energy interaction, it is impossible for the probe to read off the information about the phases $\varphi_n$ of the individual Fock states $|n\rangle$ with $n$ partons. This is because the interaction time with the proton at high energies is in general much shorter than the characteristic time scale $t_n \sim 1/E_n$ with which the Fock states evolve \cite{Kharzeev:2021nzh}. The resulting information scrambling in high-energy interactions can thus be seen as a consequence of the uncertainty relation between the phase and the occupation number of the Fock states that is similar in quantum optics\footnote{A corollary of this observation is the failure of parton model at low energies, where the scattering $S$-matrix can be parameterized in terms of phase shifts, and the experiment allows to extract the corresponding phases. The occupation number-phase uncertainty relation then implies that the occupation number cannot be determined, which makes the parton model useless at low energies.}.

The state produced from the proton after the DIS measurement is therefore a mixed state. The corresponding density matrix is obtained by tracing the density matrix of the pure proton state over the unobserved phases of the Fock states, and is characterized by non-zero von Neumann entropy. This entropy can be viewed as the entropy of entanglement between the occupation numbers and the corresponding phases in the Fock state decomposition of the proton wave function. Once this entropy is produced, it will be observed in the final state through the entropy of the multi-hadron state created in DIS. In general, the entanglement entropy may differ from the entropy of the final hadronic state. However there are several examples where the produced entanglement entropy appears equal to the entropy of the final state. One can thus hypothesize that the entanglement entropy is equal to the entropy of the hadron state, and moreover that the hadron multiplicity distribution mirrors the parton occupation number distribution inside the proton. This hypothesis can be viewed as a stronger form of the ``local parton-hadron duality". In previous studies, it has been found that this bold assumption holds surprisingly well \cite{Tu:2019ouv,Hentschinski:2021aux,Hentschinski:2022rsa,Hentschinski:2023izh}. Interestingly the model predicts that the entropy linearly grows with rapidity
what is in accord with consideration based of near conformal invariance of  
QCD in the high energy limit \cite{Lipatov:1985uk,Lipatov:1990zb, Lipatov:1993yb,Braun:2003rp} and the symmetries of the considered dipole model \cite{Gursoy:2023hge,Caputa:2024xkp}.Ref.~\cite{Dumitru:2023qee} presents an interesting calculation based on the lowest Fock states of the proton, the 3-quark state and the 3 quark+gluon one. On the other hand, our calculation, as well as the calculation of~\cite{Kharzeev:2017qzs}, assumes that the maximal entanglement takes place, as a result of a large number of partonic micro-states generated at small x due to QCD evolution. For other works on entanglement entropy at the high energy regime, see \cite{Kutak:2011rb,Peschanski:2012cw,Armesto:2019mna,Neill:2018uqw,Armesto:2019mna,Kovner:2018rbf,Chachamis:2023omp,Liu:2022hto,Liu:2023eve,Liu:2022bru,Liu:2022qqf,Stoffers:2012mn,Asadi:2023bat,Kou:2022dkw,Kutak:2023cwg,Dumitru:2023qee,Ehlers:2022oke,Ehlers:2022oal,Florio:2024aix,Grieninger:2023knz,Grieninger:2023pyb,Ikeda:2023zil,Brandenburg:2024ksp,Berges:2017hne,Berges:2017zws,Dumitru:2023fih,Dumitru:2022tud,Ramos:2022gia,Moriggi:2024tbr,Ramos:2020kaj,Dosch:2023bxj}. 

Currently employed models are able to describe successfully hadronic entropy extracted from HERA data ~\cite{H1:1996ovs,H1:2020zpd}, if  the underlying multiplicity distribution is determined inclusively, i.e., any charged hadron that hits the detector is counted. There exists however a second data set which counts hadrons only within a narrow region of rapidity along the leading parton direction in DIS, i.e. in a moving rapidity window. Unlike the global data set, which only depends on the global kinematics of the reaction, the moving rapidity window allows to study the evolution of entanglement in such a reaction, i.e. the dependence of entanglement on rapidity. 

In the small $x$ region of the DIS reactions, to which the moving rapidity model is restricted to, particle emissions are strongly ordered in rapidity, and thus in the light-cone time. The moving rapidity window therefore gives access to time evolution of entanglement entropy in the DIS reaction. To address the description of this data set, we therefore extend in the following the model used for the description of inclusive data set to the case of a local rapidity window.

Therefore, in this paper, we expand the previous studies to consider the evolution of entanglement entropy in rapidity by computing the rapidity dependence of von Neumann entropy based on QCD evolution equations. Quantitative comparison between data and our model calculations are also performed. 

{The paper is organized as follows. In Sec.~\ref{sec:theory}, the model of entanglement entropy in inclusive DIS is discussed, which includes the detail of entanglement entropy in different rapidity windows. In Sec.~\ref{sec:data}, a brief introduction on the experimental data that we compare to is given, followed by the results in Sec.~\ref{sec:numerics}. Finally, in Sec.~\ref{sec:summary}, we present a conclusion and discuss the future application of our model.}

\section{Entanglement entropy in inclusive DIS\label{sec:theory}}

\begin{figure}
\includegraphics[width=.49\textwidth]{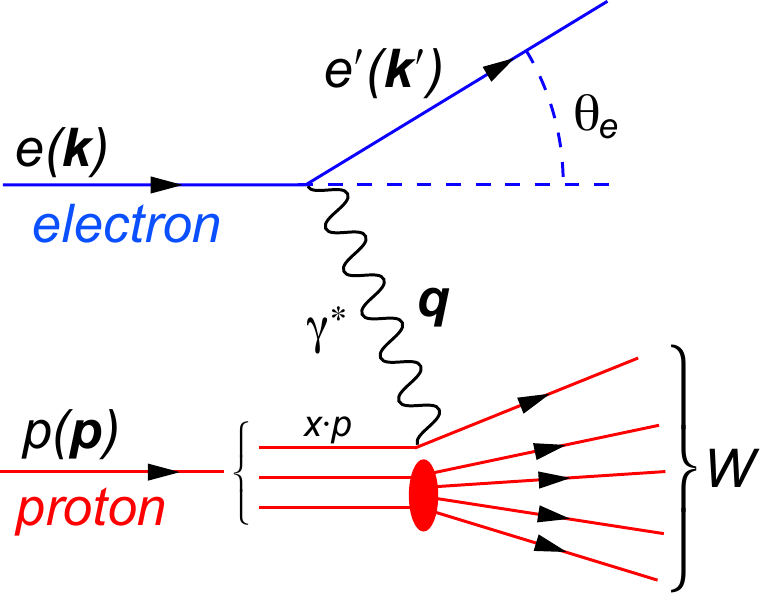}
    \caption{ Deep inelastic scattering of an electron on a proton key kinematic variables.}
    \label{fig:kinematics}
\end{figure}
We consider deep inelastic scattering of a virtual photon with momentum $q$ and virtuality $q^2 = - Q^2$ on a proton with momentum $p$ and mass $m_p$, shown in Fig.~\ref{fig:kinematics}. With the Bjorken variable  defined as $x=Q^2/(2 p \cdot q)$, the limit $x \to 1$ corresponds to elastic scattering of the photon on the proton. The low $x$ limit $x\to 0$ implies, on the other hand, a hadronic final state which is characterized by a large invariant mass of the hadronic final state, $W^2 = (1-x)Q^2/x + m_{p}^2$, and therefore, high final state multiplicities. 
A convenient framework to describe multiple particle production in the low $x$ limit is provided by the color dipole picture~\cite{Mueller:1985wy},  for a review see \cite{Kovchegov:2012mbw}. Starting with a first dipole, which arises due to the splitting of the virtual photon into a quark anti-quark pair, multiple particle production can be described within a multi-color approximation as subsequent branching of dipoles.  Due to the large relative boost factor between virtual photon and proton in the low $x$ limit,  the transverse size of  color dipoles is fixed during its interaction with the proton and the resulting $S$-matrix is diagonal in the transverse dipole size \cite{Kovchegov:2012mbw}.
Starting from the QCD evolution equation in rapidity $y = \ln 1/x$ for the color dipole amplitude \cite{Balitsky:1995ub,Kovchegov:1999ji}, it is possible to derive an evolution equation for the probability $p_n(y,\{r\})$ to encounter a state with $n$ dipoles at rapidity $y$, which takes the following generic form\cite{Mueller:1994gb,Levin:2003nc},
\begin{align}
    \label{eq:cascade1}
\partial_y p_n(y,\{r\})= \sum_m K\otimes p_m(y,\{r\}).
\end{align}
 Here $\{r\}$ denotes the  collective transverse coordinates of the $n$ dipoles. While the kernel $K$ involves in general both splitting and merging of dipole degrees of freedom, in the regime of moderate dipole densities,  which applies to our study, the evolution is dominated by splitting processes. Splitting probabilities in the kernel are then expressed in terms of kernel of the  Balitsky-Fadin-Kuraev-Lipatov (BFKL) equation \cite{Kuraev:1976ge,Kuraev:1977fs,Balitsky:1978ic}. If one further takes the limit of fixed dipole sizes during evolution, one arrives at an analytically tractable model of QCD evolution. This model of QCD evolution allows for a straightforward illustration  of  the general features of dipoles dynamics at low $x$ and  has been used both in 
\cite{Kharzeev:2017qzs} as well as the phenomenological studies \cite{Tu:2019ouv,Hentschinski:2021aux,Hentschinski:2022rsa,Hentschinski:2023izh}.  Within this setup one obtains for the probability to observe $n$ dipoles as the following result:
\begin{align}
  \label{eq:probis}
  p_n(y)& = \frac{e^{-\Delta y}}{C} \left(1-\frac{e^{-\Delta y}}{C} \right)^{n-1}.
\end{align}
Here $y = \ln 1/x$ denotes the total available phase space in rapidity, while $\Delta \simeq 0.2 - 0.4$ is the BFKL intercept of  the $1+1$ dimensional dipole model. With  $C=1$,  one fixes the initial conditions $p_1(0)=1, p_{n >1}(0)=0$. In the elastic limit $x=1$ the system consists of a single dipole and one encounters a pure state.  The mean number of dipoles is obtained as
\begin{align}
  \label{eq:mean_dipoles}
   \langle n \rangle_y  & = \sum_n  n p_n(y) = C e^{\Delta y}\equiv \bar n(x).
\end{align}
When comparing to experimental data, it is important to take into account that  only charged hadrons, predominantly charged pions, are observed in experiment. To compare to data it is necessary to rescale the mean number of dipoles by a factor $2/3$ and use $C  \to C' = 2 C/3$ in the comparison to data, see also the discussion in \cite{Hentschinski:2022rsa}. 
Using color dipoles as a suitable basis for the Schmidt decomposition in the low $x$ DIS process, one finally obtains the following expression for entanglement entropy in inclusive DIS
\begin{align}
\label{eq:inc}
    S_{inc.}(\bar{n}) & = - \sum_n p_n(\bar{n}) \ln p_n(\bar{n}) \notag \\
    &= \ln \bar{n} - (\bar{n}-1) \ln \left(1- \frac{1}{\bar{n}} \right).
\end{align}
If desired, it is straightforward, to re-express the dipole probabilities Eq.~\eqref{eq:probis} in terms of the average number of dipoles, 
\begin{align}
\label{eq:genprob}
    p_n(y) & = p_n(\bar{n}(x)), & p_n(\bar{n})&= \frac{1}{\bar{n}} \left( 1-\frac{1}{\bar{n}}\right)^{n-1}.
\end{align}
From Eq.~\eqref{eq:genprob} it is evident that in the limit of large multiplicities $\bar{n} \gg 1$, the distribution  turns into the homogeneous distribution which is characterized by maximal entanglement entropy. Indeed, one finds directly from Eq.~\eqref{eq:inc} that  $S_{inc.}(\bar{n}) =  \ln \bar{n} + 1 + {\cal O}(1/\bar{n}) \simeq  \ln \bar{n}$ for $\ln \bar{n} \gg 1$. 
Previous comparisons of entanglement entropy to  hadronic entropy extracted from H1 data,  relied on the asymptotic result $S_{inc.} = \ln (\bar{n})$
 \cite{Hentschinski:2021aux, Hentschinski:2022rsa}, where the average number of dipoles $\bar{n}$ has been directly identified with the average number of partons. Indeed the expression $S_{inc.} = \ln (\bar{n})$ appears to be more general than merely the large $\bar{n}$ limit of Eq.~\eqref{eq:inc}. While the above result has been obtained in the limit of fixed dipole sizes, Ref.~\cite{Liu:2022bru}  has solved Eq.~\eqref{eq:cascade1} allowing for variation in the dipole size. They imposed a double logarithmic approximation, where subsequent emissions are both ordered in rapidity and dipole size. While the complete Eq.~\eqref{eq:inc} could not be recovered in this limit, one also finds in this case  an entanglement entropy which takes the form $S_{inc.} = \ln \bar{n}$ for high gluon multiplicities. This observation is further supported through the large multiplicity limit of probability distributions which provide phenomenological description of charged hadron multiplicity distributions.  
 
On the other hand, the exact expression Eq.~\eqref{eq:inc} has been used for the comparison to diffractive DIS data in \cite{Hentschinski:2023izh}. In this particular case, it has been necessary to extrapolate our result to the region of intermediate $x \in [0.01, 0.5]$\footnote{Note that for diffractive studies, the corresponding parameter is usually denoted by $\beta$ instead of $x$}. One leaves the region of phase space where particle multiplicities are high and the asymptotic expression is no longer appropriate. 

 In the following we explore both versions of entanglement entropy. Since
 \begin{align}
     \label{eq:univ}
     S_{\text{inc.}}^{\text{univ.}} (\bar{n}) = \ln (\bar{n}),
 \end{align}
is not characteristic for  the probability distribution Eq.~\eqref{eq:probis}, but also arises as the large $\bar{n}$ limit of other distributions, we refer in the following to it as the ``universal entanglement entropy".
 
 We identify the average number of particles as the total number of partons 
 per $\ln 1/x$, as obtained from the sum of quark and gluon parton distribution functions (PDF):
\begin{align}
  \label{eq:mean_pdf}
\bar{n}(x, \mu) & \equiv \frac{dn}{d \ln 1/x}  = x\Sigma(x, \mu) + xg(x), \notag \\
\Sigma(x, \mu) &= \sum_f \left(q_f(x, \mu) + \bar{q}_f(x, \mu) \right).
\end{align}
Here $g(x, \mu)$ and $q_f(x,\mu)$  denote the PDF of the gluon and the quark of flavor $f$, respectively, which depend on the factorization scale which for the DIS reaction is usually identified with the photon virtuality $Q$. The $\mu$ dependence of PDFs is obtained from  the Dokshitzer-Gribov-Lipatov-Altarelli-Parisi (DGLAP) evolution equations \cite{Gribov:1972ri,Altarelli:1977zs,Dokshitzer:1977sg}. We note here that in the original K-L model \cite{Kharzeev:2017qzs}, only the gluon density was used. As demonstrated in \cite{Hentschinski:2021aux}, it is indeed the dominant contribution, but for completeness and phenomenological relevance, we also account for quarks.

As an alternative to this setup, we  further consider the complete entropy of the dipole or the Kharzeev-Levin (KL) model,
\begin{align}
    \label{eq:KL}
    S_{\text{inc.}}^{\text{KL}}(\bar{n}) & = \ln \bar{n} - (\bar{n}-1) \ln \left(1- \frac{1}{\bar{n}} \right),
\end{align}
 where we identify the average number of dipoles as 
 \begin{align}
     \label{eq:mean_KL}
     \bar{n}(x, \mu) & = \left( \frac{1}{x}\right)^{\Delta(\mu)},
 \end{align}
 with the intercept $\Delta$ obtained from a fit to parton distribution function in the low $x$ region,
 \begin{align}
     \label{eq:fit_PDF}
      C \left(\frac{1}{x}\right)^\Delta & = x\Sigma(x) + xg(x).
 \end{align}
 In both cases the mean particle number will be rescaled by a factor $2/3$ to take into account that only charged hadrons are observed in experiment. 
Numerically, both approaches are close. Indeed, $\ln C$ takes values in the range $[0.58, 1.24]$ which explains why both expressions for the dipole entropy are numerically close to each other, since the relevant difference of both entropy expressions is a relative factor of one, in the region of large hadron multiplicities. 

\begin{figure*}
    \centering\includegraphics[width=.8\textwidth]{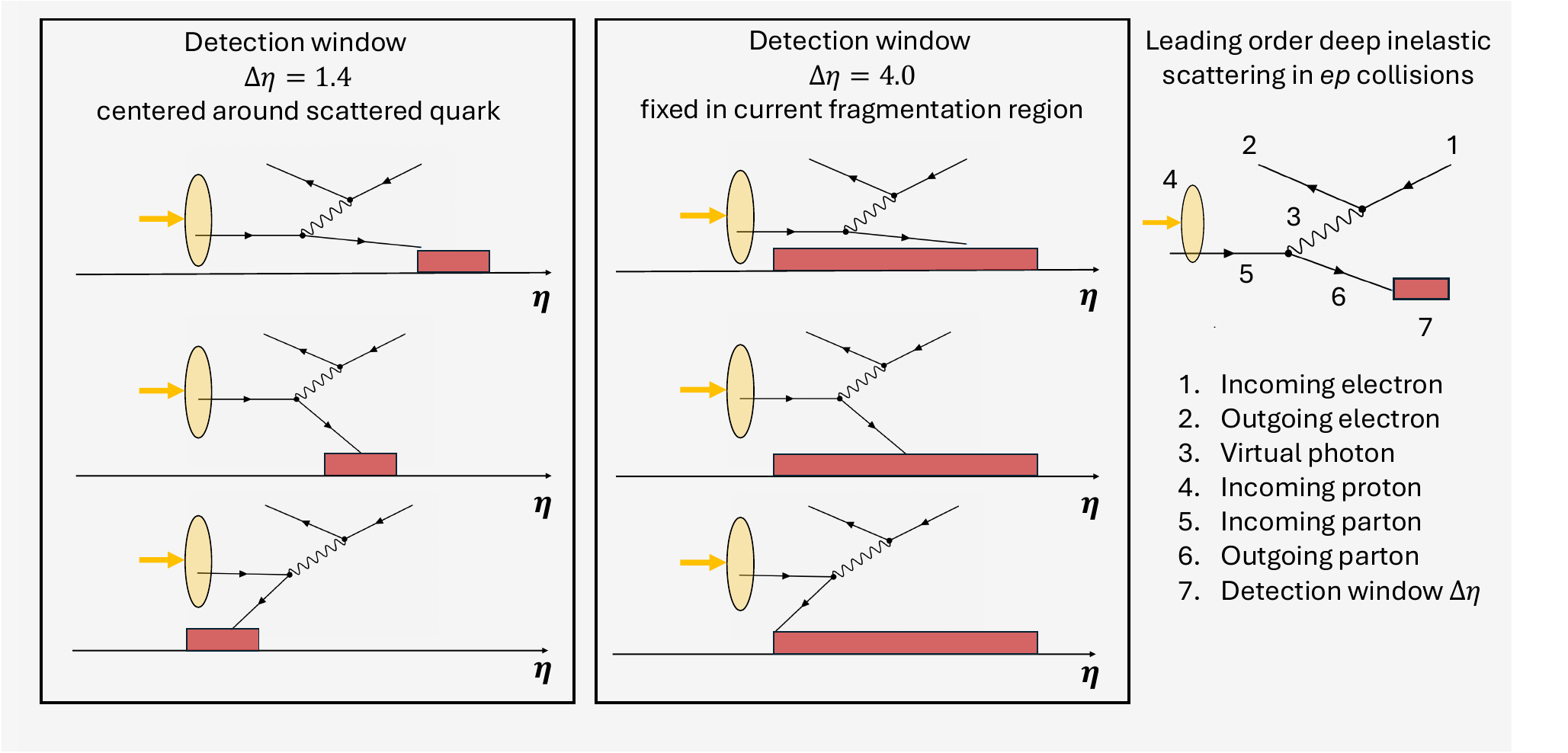}
    \caption{ Illustration of the moving detection window verse fixed detection window in the H1 measurement~\cite{H1:2020zpd}.}
    \label{fig:inc_entropies}
\end{figure*}

In the following two subsections, we extend our description to a setup where dipoles are observed in a limited region of phase space. Unlike \cite{Hentschinski:2023izh}, we do not consider the scenario, where there is no emission in a dedicated phase space region i.e. a rapidity gap. Instead we consider inclusive DIS with emissions for the entire range of rapidities, while we only observe particles in a subset. From an experimental point of view such a scenario is more appropriate for the inclusive DIS, since experiments can only access particles in a  certain sub-region in rapidity. 

\subsection{Definition of a local dipole probability distribution}

To start, we refer again to the probability distribution with fixed dipole sizes, Eq.~\eqref{eq:genprob}, with rapidity extending from $0$ up to $y = \ln 1/x$. We now divide this range into two intervals $[0, y_0]$ and $[y_0, y]$, and count only emissions in the region $[y_0, y]$. 
To address this  case, we introduce  probabilities $\tilde{p}_n(y, y_0)$ as the probability to have $n$ dipoles in the rapidity range $[y_0, y]$.  With $p_n(y_0)$ the probability to have $n$ dipoles in the range $[0, y_0]$, these new probabilities are subject to the following consistency condition: 
\begin{align}
  \label{eq:relation}
  p_n(y) & = \sum_{m=0}^n p_{n-m}(y_0) \tilde{p}_m(y, y_0).
\end{align}
The probability to have $n$ dipoles in the interval $[0, y]$ is given by all possible combinations to have $n-m$ dipoles in the range $[0, y_0]$ and $m$ in the range $[y_0, y]$. With 
\begin{align}
    \langle n \rangle_{y;y_0} & = \sum_ n n \tilde{p}_n(y, y_0),
\end{align}
and $\sum_{m=0}  \tilde{p}_m(y, y_0)=1$, this yields immediately
\begin{align}
  \label{eq:n2}
  \langle n\rangle_y & =  \langle n\rangle_{y_0} +  \langle n \rangle_{y;y_0}.
\end{align}
The average number of dipoles emitted in the interval $[0,y]$ is given as sum of the average numbers of dipoles emitted in the regions $[0, y_0]$ and $[y_0, y]$. It is straightforward to convince oneself that Eq.~\eqref{eq:relation} can be solved by 
\begin{align}
  \label{eq:ptilde}
   \tilde{p}_0(y, y_0) & =e^{-\Delta (y -y_0)} \notag \\
 \tilde{p}_{n\geq 1}(y, y_0) & = p_n(y) \cdot \left(1-\tilde{p}_0(y, y_0)   \right) .
\end{align}
Here $p_n(y)$ denote the inclusive dipole probabilities in Eq.~\eqref{eq:probis}, while $ \tilde{p}_0$ yields the probability to have no emission into the interval $[y_0, y]$.  Since $\tilde{p}_0(y,y_0)$ is the probability for no emission in the interval, $1-\tilde{p}_0(y,y_0)$ is the probability to have any number of emissions in the interval. Comparing the above probabilities to the inclusive probabilities Eq.~\eqref{eq:genprob}, it is further natural to express them as functions of the mean number of dipoles and the no-emission probability $\tp_0$,
\begin{align}
  \label{eq:ptilde_v2}
 \tilde{p}_{n\geq 1}(y, y_0) & =  \tilde{p}_{n\geq 1} (\bar{n}(x), \tp_0), \notag \\
 \tilde{p}_{n\geq 1} (\bar{n}, \tp_0) & = p_n(y) \cdot \left(1-\tilde{p}_0(y, y_0)   \right) .
\end{align}

\subsection{Definition of local entropy}

Given the above  distribution of dipoles in the region $[y_0, y]$ one finds for the entropy of the particles emitted into the interval $[y_0, y]$ the following local entropy
\begin{align}
\label{eq:slocinc}
  S_{loc}(\bar{n},\tp_0 ) & =  -\sum_{n=0} \tilde{p}_n(\bar{n}, \tp_0)\ln  \tilde{p}_n(\bar{n}, \tp_0) \notag \\
  &=
   -\tp_0 \ln \tp_0 - (1-\tp_0) \ln (1-\tp_0) \notag \\ & \hspace{1cm}+ (1-\tp_0) S_{inc.}(\bar{n}) .
\end{align}
It is given as the sum of the entropy of a two level system (emission into the rapidity interval or not) plus the  entropy of the inclusive system times the probability to have emission into the interval. Note that the above expression does not require the particular form of the probabilities of the KL-model, as shown in Eq.~\eqref{eq:cascade1}. Indeed, as long as the probabilities of Eq.~\eqref{eq:genprob}  satisfy $\sum_{n=1}p_n = 1$ together with the factorizing form Eq.~\eqref{eq:ptilde_v2}, the Eq~\eqref{eq:slocinc} will hold. 
The above expression for the local entropy fulfills the following constraints: If $\tp_0 = 1$, the probability for emission into the interval $[y_0, y]$ is zero and consequently $S_{loc} \to 0$ in this limit. For $\tp_0 = 0$, the probability distribution for emission into the intervals $[y_0, y]$ and interval $[0, y]$ coincides, and one finds $S_{loc} = S_{inc}$  as expected. 

For the following numerical studies based on Eq.~\eqref{eq:slocinc}, it will be necessary to extend our expression for the no-emission probability $\tp_0$ to 
\begin{align}
\label{eq:pot_num}
    \tp_0 &= C_0 e^{-\Delta(y-y_0)}, 
\end{align}
with a $C_0 \leq 1$ a constant factor which accounts for subleading effects that modify the no emission probability. With this parameter we will have in particular $\tp_0(y,y) = C_0 \neq 1$, i.e. the  probability for emission, $1-\tp_0$, is non-zero if the rapidity range $y-y_0$ shrinks to zero. It therefore introduces a certain non-locality in the parton-hadron duality, which relates the generated dipoles to final state hadrons. While the overall number of dipoles is fixed by the PDFs, a dipole in the range $[0, y_0]$ can generate with a certain non-zero probability a hadron in the range $[y_0, y]$. Numerically, the deviation of $C_0$ from unity is expected to be about the order of magnitude of the strong coupling constant $\alpha_s \simeq 0.2-0.3$, which characterizes the emission probability of an additional gluon. While the original dipole models with fixed dipole sizes provides the KL-model expression Eq.~\eqref{eq:KL}  as the correct expression for the inclusive entropy in Eq.~\eqref{eq:slocinc}, we will in the following also consider the case where the inclusive entropy  is determined by the universal limit Eq.~\eqref{eq:univ}.

\section{Experimental data\label{sec:data}}
The validation of our model in DIS began by comparing it to inclusive DIS data from H1~\cite{H1:2020zpd}, without considering the finite size of the rapidity window~\cite{Hentschinski:2021aux,Kharzeev:2021yyf}. We found that the contribution from sea quarks at finite energy was important although not dominant, and incorporating both gluon and quark contributions improved the agreement between the data and the model. Building on this development, we extended the model to describe another QCD process, diffractive DIS, and observed good agreement with our revised model of entanglement entropy~\cite{Hentschinski:2023izh}. This serves as an important confirmation of our model's ability to describe two different QCD processes within a consistent framework. However, as previously mentioned, an important aspect that remains to be addressed is the QCD evolution with consideration of varying rapidity windows.

Therefore, to validate the universal approach of describing the QCD evolution of entanglement entropy, including the rapidity window size dependence, the H1 $ep$ DIS data~\cite{H1:2020zpd} from HERA was used. There are two distinct measurements of entanglement entropy in the H1 experiment, see Fig.~\ref{fig:inc_entropies} for an illustration:
\begin{itemize}
    \item A narrow detection window centered around the scattered quark with $\Delta\eta=1.4$;
    \item A fixed detection window in the current fragmentation region with $\Delta\eta^{\ast}=4.0$\footnote{$\eta^{\ast}$is the pseudorapidity variable in the hadronic center-of-mass frame.} ($\Delta\eta\approx 3.2$).
\end{itemize}

In the leading DIS picture, the leading quark can be estimated from the kinematic variables $x$ and $Q^{2}$~\cite{Tu:2019ouv}, and the pseudorapidity ranges are corresponding to the respective $x$ range. 

Intuitively, the leading scattered quark goes more forward for high $x$ events (top left in Fig.~\ref{fig:inc_entropies}) as high $x$ quark has a larger momentum fraction of the incoming proton that has a smaller scattering angle. The detection window of 1.4 unit of pseudorapidity is placed, therefore, more forward. As the energy of the virtual photon increases at the same virtuality $Q^2$, the $x$ decreases and the scattering angle of the leading quark increases. Thus, the detection window moves backward, as illustrated in the next two sub-panels on the left in Fig.~\ref{fig:inc_entropies}.

For the fixed detection window, the position of the window is chosen to be qualitatively the current fragmentation region. Specifically, the DIS events are boosted into the hadronic center-of-mass (HCM) frame~\cite{H1:2020zpd} and the photon-going direction is defined as the positive pseudorapidity. The current fragmentation region is then selected as the pseudorapidity in the HCM frame $0<\eta^{\ast}<4$, which has to be within $-1.6<\eta_{lab}<1.6$. Note that the lab $\eta$ selection is the boundary of the tracking acceptance of the H1 detector. Most of the current fragmentation region overlaps with this lab $\eta$ selection. This is shown in Fig.~\ref{fig:inc_entropies} on the right. 

Future studies of other QCD processes are also of great interest. For instance, will maximal entanglement be achieved in hard QCD scattering processes, such as events involving high transverse momentum jets and their fragmentation? Could heavy flavor productions offer additional insights into the entanglement-based approach for understanding the parton structure? These questions could be highly relevant for future experimental investigations.

\section{Numerical results}
\label{sec:numerics}

\begin{figure*}
    \centering     \includegraphics[width=.9\textwidth]{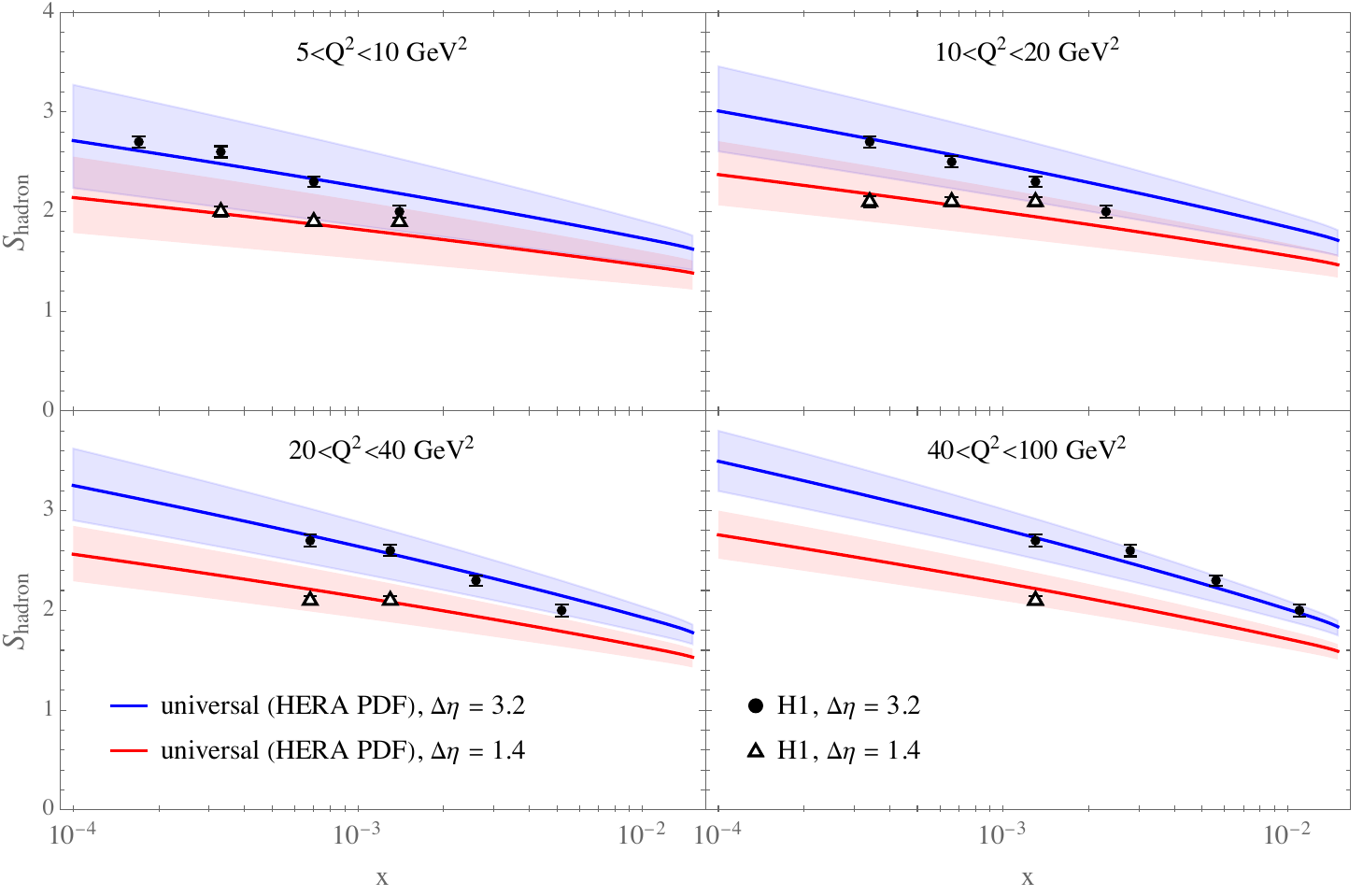}

    \caption{Hadronic entropy as obtained from Eq.~\eqref{eq:slocinc} employing Eq.~\eqref{eq:univ} for the inclusive entropy with the number of dipoles determined through Eq.~\eqref{eq:mean_pdf} based on leading order HERAPDFs. The no-emission probability has been determined from Eq.~\eqref{eq:pot_num} using the $Delta$ parameter extracted from HERAPDFs, see also the appendix Sec.~\ref{sec:fitdetails} and the corresponding parameter $C_0$ presented in Tab.~\ref{tab:C0} }
    \label{fig:loc}
\end{figure*}

In previous comparisons to experiment, data collected with a fixed detection window have been referred to as inclusive data \cite{Hentschinski:2021aux,Hentschinski:2022rsa,Hentschinski:2023izh}. On the other hand, the availability of two data sets allows us to further test the robustness of the approach developed in Sec.~\ref{sec:theory}. We will treat in the following both data sets with their regarding limitation in the range in which data have been recorded, i.e.,  $\Delta \eta \approx y - y_0 = 3.2$ and  $\Delta \eta \approx y - y_0 = 1.4$.

 To arrive at our theory predictions, we use the following procedure: 
    \begin{itemize}
        \item     We first determine the  intercept $\Delta$, together with the constant $C$, from a fit to leading order PDFs, using the relations Eq.~\eqref{eq:fit_PDF} in the low $x$ limit, $x \in [10^{-3}, 10^{-5}]$, where the power law Eq.~\eqref{eq:fit_PDF} holds to a good approximation. For the parton distributions we use leading order HERAPDFs \cite{H1:2015ubc} and PDFs obtained from the HSS unintegrated gluon distribution \cite{Hentschinski:2012kr,Hentschinski:2013id,Chachamis:2015ona,Bautista:2016xnp}, which is subject to next-to-leading order BFKL evolution. For both PDF sets we identify the hard scale with the photon virtuality $Q=\mu$. For technical details see the discussion in the Appendix \ref{sec:fitdetails}. For further details on PDFs extracted from the HSS fit, see \cite{Hentschinski:2021aux,Hentschinski:2022rsa}. 
        
        \item To estimate uncertainties on the extracted  parameters we vary the hard scale $\mu$ of the PDFs in the interval $\mu \in [Q/2, 2Q]$. We further vary the range in $x$ over which the fits are performed, i.e., we take the ranges   $x \in [10^{-2}, 10^{-4}]$ and $x \in [10^{-4}, 10^{-6}]$. Uncertainties due to the choice of the hard scale and uncertainties due to a variation in the range of $x$  are added in quadrature to yield the complete uncertainty. For HERAPDFs, we further include the experimental uncertainty; for HSS PDFs this information is not available. The results of these fits are summarized in Tab.~\ref{tab:Delta_C} of the Appendix.
        
        \item Using the extracted intercept $\Delta$,  we then determine the parameter $C_0$, which controls the local no-emission probability from a global fit  of Eq.~\eqref{eq:slocinc} to the H1 data. For this fit we include  the four available bins in photon virtuality $Q^2$ and consider data for both  $\Delta \eta = 3.2$  and $\Delta \eta = 1.4$. For the inclusive entropy $S_{inc.}$ in Eq.~\eqref{eq:slocinc}  we use both the expression obtained from the KL-model Eq.~\eqref{eq:KL} as well  the universal entropy of Eq.~\eqref{eq:univ}. 
        The results of the fit of the $C_0$ parameter are presented in Tab.~\ref{tab:C0}. 
    \end{itemize}
As a result of our fit of the $C_0$ parameters, we find that the data set prefers a $C_0$ of roughly 0.7, i.e. the emission probability at $y=y_0$ is of the order of $0.3$, which is in accordance with our expectations. We finally present in Tab.~\ref{tab:chi2} the values of the $\chi^2$/d.o.f for each of the fits. 
\begin{table}[bt]
    \centering
     \caption{Values of the constant $C_0$ obtained from a fit to data.  The uncertainties are directly related to the uncertainties of the underlying data set.   }
    \begin{tabular}{l|c|c|c}
    $C_0$  &   HSS PDFs& LO HERAPDF   \\ \hline \hline
    KL model   &  $0.655 \pm 0.009$  & $0.777 \pm 0.008$  \\ \hline
  %  universal &  $0.710 \pm 0.008$ & $0.797 \pm 0.007$ \\ \hline
     universal  &  $0.777 \pm 0.008$ & $0.706 \pm 0.008$ 
    \end{tabular}\\

    \label{tab:C0}
\end{table}
\begin{table}[bt]
    \centering
    \caption{$\chi^2/$d.o.f of each description. In brackets with the uncertainty including both experimental and theory uncertainty, added in quadrature.}
    
    \begin{tabular}{l|c|c|c}
     $\chi^2/$d.o.f &   HSS PDFs& LO HERAPDF   \\ \hline
    KL model   & 10.6  (0.67) &  4.5 (0.08)\\ \hline
 %   universal &4.7 (0.23) &  6.1 (0.10) \\ \hline
    universal & 4.35 (0.21) & 4.18 (0.22)
    \end{tabular}
    
    \label{tab:chi2}
\end{table}
While the fit includes only the experimental errors, we include in Tab.~\ref{tab:chi2} also the corresponding values with  experimental and theoretical error combined. While the quality of the description differs for different underlying theoretical models, we find for all models an excellent description of data, once both experimental and theoretical uncertainties are accounted for. 

We finally show comparison of our theory predictions to experimental data  based on HERAPDF using the  universal entropy formula in Fig.~\ref{fig:loc}. We find that both the local ($\Delta \eta = 1.4$) and inclusive ($\Delta \eta = 3.2$) are very well described by our model. We, therefore, confirm that our generalized formalism can indeed account for the emission into a local rapidity window and describe both data taken for a small and large rapidity window.

\section{Conclusion~\label{sec:summary}}
In conclusion, we have studied the QCD evolution of entanglement entropy of the proton in DIS for the first time. We generalized the formalism by Kharzeev and Levin \cite{Kharzeev:2017qzs} to scenarios where final state hadrons are produced in different rapidity intervals. Specifically, our generalized model describes data from both a fixed rapidity window that covers most of the current fragmentation region and a moving, narrow rapidity window centered around the leading parton in DIS. The significant variation in the rapidity intervals of these measurements is sensitive to the QCD evolution of parton distributions. The new formalism does not require the specific for original K-L model form of the probabilities $P_n$. By using different form of $P_n$ it can be used to address saturation effects and dipole evolution in a 3+1 D scenario \cite{Kharzeev:2017qzs}.

Our approach generalizes the original Kharzeev-Levin model to the case of entanglement entropy produced in a specific rapidity interval, and successfully describes all available data simultaneously. Implementing QCD evolution of entanglement entropy marks a key advancement in refining this approach. Future applications could explore entanglement within a jet that 
may provide new insights into fragmentation and hadronization, modifications of parton distributions in nuclei, and the nonperturbative regime of parton distributions, thereby offering a new theoretical and experimental perspective on QCD confinement.

\newpage
\section*{Acknowledgements} 

M.~Hentschinski acknowledges support by UDLAP Apoyos VAC 2024 and is grateful for hospitality at the INP Krak\'ow. The work of D.~Kharzeev is supported by the U.S. Department of Energy, Office of Science, Office of Nuclear Physics, Grants No. DE-FG88ER41450 and DE-SC0012704 and by the U.S. Department of Energy, Office of Science, National Quantum Information Science Research Centers, Co-design Center for Quantum Advantage (C2QA) under Contract No.DE-SC0012704. The work of K.~ Kutak has been partially supported by  European Union’s Horizon 2020  research  and  innovation  program  under  grant  agreement  No.824093. The work of Z.~Tu is supported by the U.S. Department of Energy under Award DE-SC0012704, IANN-QCD award 2024, and the BNL Laboratory Directed Research and Developement (LDRD) 23-050 project. Z.~Tu thanks UDLAP in Mexico for their hospitality during the period this project was being worked out.  

\appendix
\begin{table*}[tb]
   % \centering
     \caption{Paramters $C, \Delta$ extracted from a comparison to HSS and HERA PDF. }
    \begin{tabular}{l|c|c|c|c}
        $Q^2/$GeV$^2 \in$ &  [5, 10]  & [10,20] & [20, 40] & [40, 100]  \\
        \hline \hline 
       $\Delta_{\text{HSS}}$   &  0.27 $\pm$ 0.04& 0.28 $\pm$ 0.04 & 0.29 $\pm$ 0.05 & 0.29 $\pm$ 0.04
       \\
         $\Delta_{\text{HERAPDF}}$   &  0.28 $\pm$ 0.08 &
         0.31 $\pm$ 0.07 & 0.33 $\pm$ 0.06 & 0.36 $\pm$ 0.06
       \\
       $C_{\text{HSS}}$ & 2.1 $\pm$ 0.3 & 2.7 $\pm$ 0.4 & 3.2 $\pm$ 0.5 & 3.9 $\pm$ 0.7 \\
       $C_{\text{HERAPDF}}$ & 2.54 $\pm$ 0.68 & 2.62 $\pm$ 0.78 & 2.65 $\pm$ 0.91 & 2.66 $\pm$ 1.02
    \end{tabular}
   
    \label{tab:Delta_C}
\end{table*}  

\section{Details on the fit of the parameters $\Delta, C$\label{sec:appendix-a}}
\label{sec:fitdetails}
 To obtain the parameter $\Delta$ for the determination of the the no-emission probability as well as for the evaluation of the KL model, we first average  the parton distributions over bins in photon virtuality $Q^2$,
 \begin{align}
         &   \left \langle \frac{dn}{d \ln 1/x}  \right \rangle  = \frac{1}{Q^2_{max} - Q^2_{min}} \notag \\
          & \hspace{2cm} \int_{Q^2_{min}}^{Q^2_{max}} dQ^2 \left( xg(x, \mu) + x\Sigma(x, \mu) \right).
        \end{align}
We then fit the relation Eq.~\eqref{eq:fit_PDF} for each bin in rapidity and obtain in this way for each bin a set of parameters $\Delta, C$. The dependence of these parameters on $Q^2$ is then a direct consequence of leading order DGLAP evolution \cite{Gribov:1972ri,Altarelli:1977zs,Dokshitzer:1977sg} in the case of HERAPDFs and next-to-leading order BFKL evolution \cite{Fadin:1998py}, combined with a resummation of collinear logarithms inside the next-to-leading order BFKL kernel. Theory uncertainties of the HSS gluon include a variation of the scale of the running coupling as introduced in \cite{Bautista:2016xnp}. For the determination of PDFs from the HSS unintegrated gluon density see \cite{Hentschinski:2021aux,Hentschinski:2022rsa}. To evaluate HERAPDFs, the ManePare package \cite{Clark:2016jgm} has been used.

The results of these fits are summarized in Tab.~\ref{tab:Delta_C}. While the intercept $\Delta$ extracted from  leading order HERAPDFs is in general higher than the one obtained from HSS PDFs,  both values agree within uncertainties. A similar statement applies to the constant $C$. We however observe a considerable $Q^2$ dependence of the $C$ in the case of HSS PDFs, whereas the same parameter grows only weakly in the case of HERAPDFs. 

\section{Universal vs. KL model\label{sec:appendix-b}}
In this paragraph we examine the quality of the description of the universal and KL model entropy of the data set with $\Delta \eta = 3.2$, if the theory description does not take into account this restriction onto the phase space, i.e. Eq.~\eqref{eq:KL} and Eq.~\eqref{eq:univ} are used directly to obtain theory predictions. Previous phenomenological studies \cite{Hentschinski:2022rsa,Hentschinski:2021aux} applied only the universal expressions and it is therefore of interest to examine to which extend the KL model can provide a satisfactory description in this case. 
\begin{table}[tb]
    \centering
      \caption{Comparison of the inclusive expressions to data. In brackets the $\chi^2/$dof. if both experimental and theory uncertainties are included.}
    \begin{tabular}{c|c|c} 
       $\chi^2$/dof.  &  HSS PDF & HERA PDF\\ \hline \hline 
        KL model &  16.2 (0.89) & 14.1 (0.50) \\ \hline
        universal & 6.95 (0.11) & 14.9 (0.39)
    \end{tabular}
  
    \label{tab:inclusive}
\end{table}
To access the quality of the description, we determine $\chi^2$ per degree of freedom. Our results are shown in Tab.~\ref{tab:inclusive}. While the universal description, which uses the complete PDF provides in general a better description of data, also the KL model gives a very good description, in particular if theory uncertainties are included. If the KL model is evaluated using the HSS value for the intercept $\Delta$, we find that $Q^2$ evolution of this parameter is insufficient to describe also the highest $Q^2$ bins. On the other hand, universal entropy provides the best description, if the number of partons per unit $\ln 1/x$ is determined from HSS PDFs. HERAPDFs provide a description of similar quality, for both KL and universal entropy.

\bibliography{paper}

%merlin.mbs apsrev4-1.bst 2010-07-25 4.21a (PWD, AO, DPC) hacked
%Control: key (0)
%Control: author (8) initials jnrlst
%Control: editor formatted (1) identically to author
%Control: production of article title (-1) disabled
%Control: page (0) single
%Control: year (1) truncated
%Control: production of eprint (0) enabled
\begin{thebibliography}{79}%
\makeatletter
\providecommand \@ifxundefined [1]{%
 \@ifx{#1\undefined}
}%
\providecommand \@ifnum [1]{%
 \ifnum #1\expandafter \@firstoftwo
 \else \expandafter \@secondoftwo
 \fi
}%
\providecommand \@ifx [1]{%
 \ifx #1\expandafter \@firstoftwo
 \else \expandafter \@secondoftwo
 \fi
}%
\providecommand \natexlab [1]{#1}%
\providecommand \enquote  [1]{``#1''}%
\providecommand \bibnamefont  [1]{#1}%
\providecommand \bibfnamefont [1]{#1}%
\providecommand \citenamefont [1]{#1}%
\providecommand \href@noop [0]{\@secondoftwo}%
\providecommand \href [0]{\begingroup \@sanitize@url \@href}%
\providecommand \@href[1]{\@@startlink{#1}\@@href}%
\providecommand \@@href[1]{\endgroup#1\@@endlink}%
\providecommand \@sanitize@url [0]{\catcode `\\12\catcode `\$12\catcode `\&12\catcode `\#12\catcode `\^12\catcode `\_12\catcode `\%12\relax}%
\providecommand \@@startlink[1]{}%
\providecommand \@@endlink[0]{}%
\providecommand \url  [0]{\begingroup\@sanitize@url \@url }%
\providecommand \@url [1]{\endgroup\@href {#1}{\urlprefix }}%
\providecommand \urlprefix  [0]{URL }%
\providecommand \Eprint [0]{\href }%
\providecommand \doibase [0]{http://dx.doi.org/}%
\providecommand \selectlanguage [0]{\@gobble}%
\providecommand \bibinfo  [0]{\@secondoftwo}%
\providecommand \bibfield  [0]{\@secondoftwo}%
\providecommand \translation [1]{[#1]}%
\providecommand \BibitemOpen [0]{}%
\providecommand \bibitemStop [0]{}%
\providecommand \bibitemNoStop [0]{.\EOS\space}%
\providecommand \EOS [0]{\spacefactor3000\relax}%
\providecommand \BibitemShut  [1]{\csname bibitem#1\endcsname}%
\let\auto@bib@innerbib\@empty
%</preamble>
\bibitem [{\citenamefont {Kharzeev}\ and\ \citenamefont {Levin}(2017)}]{Kharzeev:2017qzs}%
  \BibitemOpen
  \bibfield  {author} {\bibinfo {author} {\bibfnamefont {D.~E.}\ \bibnamefont {Kharzeev}}\ and\ \bibinfo {author} {\bibfnamefont {E.~M.}\ \bibnamefont {Levin}},\ }\href {\doibase 10.1103/PhysRevD.95.114008} {\bibfield  {journal} {\bibinfo  {journal} {Phys. Rev. D}\ }\textbf {\bibinfo {volume} {95}},\ \bibinfo {pages} {114008} (\bibinfo {year} {2017})},\ \Eprint {http://arxiv.org/abs/1702.03489} {arXiv:1702.03489 [hep-ph]} \BibitemShut {NoStop}%
\bibitem [{\citenamefont {Alba}\ and\ \citenamefont {Calabrese}(2017)}]{Alba:2017ekd}%
  \BibitemOpen
  \bibfield  {author} {\bibinfo {author} {\bibfnamefont {V.}~\bibnamefont {Alba}}\ and\ \bibinfo {author} {\bibfnamefont {P.}~\bibnamefont {Calabrese}},\ }\href {\doibase 10.1073/pnas.1703516114} {\bibfield  {journal} {\bibinfo  {journal} {Proc. Nat. Acad. Sci.}\ }\textbf {\bibinfo {volume} {114}},\ \bibinfo {pages} {7947} (\bibinfo {year} {2017})}\BibitemShut {NoStop}%
\bibitem [{\citenamefont {Berges}\ \emph {et~al.}(2021)\citenamefont {Berges}, \citenamefont {Heller}, \citenamefont {Mazeliauskas},\ and\ \citenamefont {Venugopalan}}]{Berges:2020fwq}%
  \BibitemOpen
  \bibfield  {author} {\bibinfo {author} {\bibfnamefont {J.}~\bibnamefont {Berges}}, \bibinfo {author} {\bibfnamefont {M.~P.}\ \bibnamefont {Heller}}, \bibinfo {author} {\bibfnamefont {A.}~\bibnamefont {Mazeliauskas}}, \ and\ \bibinfo {author} {\bibfnamefont {R.}~\bibnamefont {Venugopalan}},\ }\href {\doibase 10.1103/RevModPhys.93.035003} {\bibfield  {journal} {\bibinfo  {journal} {Rev. Mod. Phys.}\ }\textbf {\bibinfo {volume} {93}},\ \bibinfo {pages} {035003} (\bibinfo {year} {2021})},\ \Eprint {http://arxiv.org/abs/2005.12299} {arXiv:2005.12299 [hep-th]} \BibitemShut {NoStop}%
\bibitem [{\citenamefont {Bhattacharya}\ \emph {et~al.}(2024)\citenamefont {Bhattacharya}, \citenamefont {Boussarie},\ and\ \citenamefont {Hatta}}]{Bhattacharya:2024sno}%
  \BibitemOpen
  \bibfield  {author} {\bibinfo {author} {\bibfnamefont {S.}~\bibnamefont {Bhattacharya}}, \bibinfo {author} {\bibfnamefont {R.}~\bibnamefont {Boussarie}}, \ and\ \bibinfo {author} {\bibfnamefont {Y.}~\bibnamefont {Hatta}},\ }\href@noop {} {\  (\bibinfo {year} {2024})},\ \Eprint {http://arxiv.org/abs/2404.04208} {arXiv:2404.04208 [hep-ph]} \BibitemShut {NoStop}%
\bibitem [{\citenamefont {Gong}\ \emph {et~al.}(2022)\citenamefont {Gong}, \citenamefont {Parida}, \citenamefont {Tu},\ and\ \citenamefont {Venugopalan}}]{Gong:2021bcp}%
  \BibitemOpen
  \bibfield  {author} {\bibinfo {author} {\bibfnamefont {W.}~\bibnamefont {Gong}}, \bibinfo {author} {\bibfnamefont {G.}~\bibnamefont {Parida}}, \bibinfo {author} {\bibfnamefont {Z.}~\bibnamefont {Tu}}, \ and\ \bibinfo {author} {\bibfnamefont {R.}~\bibnamefont {Venugopalan}},\ }\href {\doibase 10.1103/PhysRevD.106.L031501} {\bibfield  {journal} {\bibinfo  {journal} {Phys. Rev. D}\ }\textbf {\bibinfo {volume} {106}},\ \bibinfo {pages} {L031501} (\bibinfo {year} {2022})},\ \Eprint {http://arxiv.org/abs/2107.13007} {arXiv:2107.13007 [hep-ph]} \BibitemShut {NoStop}%
\bibitem [{\citenamefont {Aoude}\ \emph {et~al.}(2022)\citenamefont {Aoude}, \citenamefont {Madge}, \citenamefont {Maltoni},\ and\ \citenamefont {Mantani}}]{Aoude:2022imd}%
  \BibitemOpen
  \bibfield  {author} {\bibinfo {author} {\bibfnamefont {R.}~\bibnamefont {Aoude}}, \bibinfo {author} {\bibfnamefont {E.}~\bibnamefont {Madge}}, \bibinfo {author} {\bibfnamefont {F.}~\bibnamefont {Maltoni}}, \ and\ \bibinfo {author} {\bibfnamefont {L.}~\bibnamefont {Mantani}},\ }\href {\doibase 10.1103/PhysRevD.106.055007} {\bibfield  {journal} {\bibinfo  {journal} {Phys. Rev. D}\ }\textbf {\bibinfo {volume} {106}},\ \bibinfo {pages} {055007} (\bibinfo {year} {2022})},\ \Eprint {http://arxiv.org/abs/2203.05619} {arXiv:2203.05619 [hep-ph]} \BibitemShut {NoStop}%
\bibitem [{\citenamefont {Severi}\ \emph {et~al.}(2022)\citenamefont {Severi}, \citenamefont {Boschi}, \citenamefont {Maltoni},\ and\ \citenamefont {Sioli}}]{Severi:2021cnj}%
  \BibitemOpen
  \bibfield  {author} {\bibinfo {author} {\bibfnamefont {C.}~\bibnamefont {Severi}}, \bibinfo {author} {\bibfnamefont {C.~D.~E.}\ \bibnamefont {Boschi}}, \bibinfo {author} {\bibfnamefont {F.}~\bibnamefont {Maltoni}}, \ and\ \bibinfo {author} {\bibfnamefont {M.}~\bibnamefont {Sioli}},\ }\href {\doibase 10.1140/epjc/s10052-022-10245-9} {\bibfield  {journal} {\bibinfo  {journal} {Eur. Phys. J. C}\ }\textbf {\bibinfo {volume} {82}},\ \bibinfo {pages} {285} (\bibinfo {year} {2022})},\ \Eprint {http://arxiv.org/abs/2110.10112} {arXiv:2110.10112 [hep-ph]} \BibitemShut {NoStop}%
\bibitem [{\citenamefont {White}\ and\ \citenamefont {White}(2024)}]{White:2024nuc}%
  \BibitemOpen
  \bibfield  {author} {\bibinfo {author} {\bibfnamefont {C.~D.}\ \bibnamefont {White}}\ and\ \bibinfo {author} {\bibfnamefont {M.~J.}\ \bibnamefont {White}},\ }\href@noop {} {\  (\bibinfo {year} {2024})},\ \Eprint {http://arxiv.org/abs/2406.07321} {arXiv:2406.07321 [hep-ph]} \BibitemShut {NoStop}%
\bibitem [{CMS(2024)}]{CMS:2024pts}%
  \BibitemOpen
  \href@noop {} {\  (\bibinfo {year} {2024})},\ \Eprint {http://arxiv.org/abs/2406.03976} {arXiv:2406.03976 [hep-ex]} \BibitemShut {NoStop}%
\bibitem [{\citenamefont {Aad}\ \emph {et~al.}(2023)\citenamefont {Aad} \emph {et~al.}}]{ATLAS:2023fsd}%
  \BibitemOpen
  \bibfield  {author} {\bibinfo {author} {\bibfnamefont {G.}~\bibnamefont {Aad}} \emph {et~al.} (\bibinfo {collaboration} {ATLAS}),\ }\href@noop {} {\  (\bibinfo {year} {2023})},\ \Eprint {http://arxiv.org/abs/2311.07288} {arXiv:2311.07288 [hep-ex]} \BibitemShut {NoStop}%
\bibitem [{\citenamefont {Banuls}\ \emph {et~al.}(2023)\citenamefont {Banuls}, \citenamefont {Heller}, \citenamefont {Jansen}, \citenamefont {Knaute},\ and\ \citenamefont {Svensson}}]{Banuls:2022iwk}%
  \BibitemOpen
  \bibfield  {author} {\bibinfo {author} {\bibfnamefont {M.~C.}\ \bibnamefont {Banuls}}, \bibinfo {author} {\bibfnamefont {M.~P.}\ \bibnamefont {Heller}}, \bibinfo {author} {\bibfnamefont {K.}~\bibnamefont {Jansen}}, \bibinfo {author} {\bibfnamefont {J.}~\bibnamefont {Knaute}}, \ and\ \bibinfo {author} {\bibfnamefont {V.}~\bibnamefont {Svensson}},\ }\href {\doibase 10.1103/PhysRevD.108.076016} {\bibfield  {journal} {\bibinfo  {journal} {Phys. Rev. D}\ }\textbf {\bibinfo {volume} {108}},\ \bibinfo {pages} {076016} (\bibinfo {year} {2023})},\ \Eprint {http://arxiv.org/abs/2206.10528} {arXiv:2206.10528 [hep-th]} \BibitemShut {NoStop}%
\bibitem [{\citenamefont {Banerjee}\ \emph {et~al.}(2005)\citenamefont {Banerjee}, \citenamefont {Ghosh}, \citenamefont {Ilgenfritz}, \citenamefont {Raha}, \citenamefont {Takasugi},\ and\ \citenamefont {Toki}}]{Banerjee:2003fg}%
  \BibitemOpen
  \bibfield  {author} {\bibinfo {author} {\bibfnamefont {S.}~\bibnamefont {Banerjee}}, \bibinfo {author} {\bibfnamefont {S.~K.}\ \bibnamefont {Ghosh}}, \bibinfo {author} {\bibfnamefont {E.-M.}\ \bibnamefont {Ilgenfritz}}, \bibinfo {author} {\bibfnamefont {S.}~\bibnamefont {Raha}}, \bibinfo {author} {\bibfnamefont {E.}~\bibnamefont {Takasugi}}, \ and\ \bibinfo {author} {\bibfnamefont {H.}~\bibnamefont {Toki}},\ }\href {\doibase 10.1016/j.physletb.2005.02.008} {\bibfield  {journal} {\bibinfo  {journal} {Phys. Lett. B}\ }\textbf {\bibinfo {volume} {611}},\ \bibinfo {pages} {27} (\bibinfo {year} {2005})},\ \Eprint {http://arxiv.org/abs/hep-ph/0307366} {arXiv:hep-ph/0307366} \BibitemShut {NoStop}%
\bibitem [{\citenamefont {Klebanov}\ \emph {et~al.}(2008)\citenamefont {Klebanov}, \citenamefont {Kutasov},\ and\ \citenamefont {Murugan}}]{Klebanov:2007ws}%
  \BibitemOpen
  \bibfield  {author} {\bibinfo {author} {\bibfnamefont {I.~R.}\ \bibnamefont {Klebanov}}, \bibinfo {author} {\bibfnamefont {D.}~\bibnamefont {Kutasov}}, \ and\ \bibinfo {author} {\bibfnamefont {A.}~\bibnamefont {Murugan}},\ }\href {\doibase 10.1016/j.nuclphysb.2007.12.017} {\bibfield  {journal} {\bibinfo  {journal} {Nucl. Phys. B}\ }\textbf {\bibinfo {volume} {796}},\ \bibinfo {pages} {274} (\bibinfo {year} {2008})},\ \Eprint {http://arxiv.org/abs/0709.2140} {arXiv:0709.2140 [hep-th]} \BibitemShut {NoStop}%
\bibitem [{\citenamefont {Grieninger}\ \emph {et~al.}(2023{\natexlab{a}})\citenamefont {Grieninger}, \citenamefont {Kharzeev},\ and\ \citenamefont {Zahed}}]{Grieninger:2023ehb}%
  \BibitemOpen
  \bibfield  {author} {\bibinfo {author} {\bibfnamefont {S.}~\bibnamefont {Grieninger}}, \bibinfo {author} {\bibfnamefont {D.~E.}\ \bibnamefont {Kharzeev}}, \ and\ \bibinfo {author} {\bibfnamefont {I.}~\bibnamefont {Zahed}},\ }\href {\doibase 10.1103/PhysRevD.108.086030} {\bibfield  {journal} {\bibinfo  {journal} {Phys. Rev. D}\ }\textbf {\bibinfo {volume} {108}},\ \bibinfo {pages} {086030} (\bibinfo {year} {2023}{\natexlab{a}})},\ \Eprint {http://arxiv.org/abs/2305.07121} {arXiv:2305.07121 [hep-th]} \BibitemShut {NoStop}%
\bibitem [{\citenamefont {Nielsen}\ and\ \citenamefont {Chuang}(2012)}]{Nielsen:2012yss}%
  \BibitemOpen
  \bibfield  {author} {\bibinfo {author} {\bibfnamefont {M.~A.}\ \bibnamefont {Nielsen}}\ and\ \bibinfo {author} {\bibfnamefont {I.~L.}\ \bibnamefont {Chuang}},\ }\href {\doibase 10.1017/cbo9780511976667} {\emph {\bibinfo {title} {{Quantum Computation and Quantum Information}}}}\ (\bibinfo  {publisher} {Cambridge University Press},\ \bibinfo {year} {2012})\BibitemShut {NoStop}%
\bibitem [{\citenamefont {Horodecki}\ \emph {et~al.}(2009)\citenamefont {Horodecki}, \citenamefont {Horodecki}, \citenamefont {Horodecki},\ and\ \citenamefont {Horodecki}}]{Horodecki:2009zz}%
  \BibitemOpen
  \bibfield  {author} {\bibinfo {author} {\bibfnamefont {R.}~\bibnamefont {Horodecki}}, \bibinfo {author} {\bibfnamefont {P.}~\bibnamefont {Horodecki}}, \bibinfo {author} {\bibfnamefont {M.}~\bibnamefont {Horodecki}}, \ and\ \bibinfo {author} {\bibfnamefont {K.}~\bibnamefont {Horodecki}},\ }\href {\doibase 10.1103/RevModPhys.81.865} {\bibfield  {journal} {\bibinfo  {journal} {Rev. Mod. Phys.}\ }\textbf {\bibinfo {volume} {81}},\ \bibinfo {pages} {865} (\bibinfo {year} {2009})},\ \Eprint {http://arxiv.org/abs/quant-ph/0702225} {arXiv:quant-ph/0702225} \BibitemShut {NoStop}%
\bibitem [{\citenamefont {Kharzeev}(2021)}]{Kharzeev:2021nzh}%
  \BibitemOpen
  \bibfield  {author} {\bibinfo {author} {\bibfnamefont {D.~E.}\ \bibnamefont {Kharzeev}},\ }\href {\doibase 10.1098/rsta.2021.0063} {\bibfield  {journal} {\bibinfo  {journal} {Phil. Trans. A. Math. Phys. Eng. Sci.}\ }\textbf {\bibinfo {volume} {380}},\ \bibinfo {pages} {20210063} (\bibinfo {year} {2021})},\ \Eprint {http://arxiv.org/abs/2108.08792} {arXiv:2108.08792 [hep-ph]} \BibitemShut {NoStop}%
\bibitem [{\citenamefont {Tu}\ \emph {et~al.}(2020)\citenamefont {Tu}, \citenamefont {Kharzeev},\ and\ \citenamefont {Ullrich}}]{Tu:2019ouv}%
  \BibitemOpen
  \bibfield  {author} {\bibinfo {author} {\bibfnamefont {Z.}~\bibnamefont {Tu}}, \bibinfo {author} {\bibfnamefont {D.~E.}\ \bibnamefont {Kharzeev}}, \ and\ \bibinfo {author} {\bibfnamefont {T.}~\bibnamefont {Ullrich}},\ }\href {\doibase 10.1103/PhysRevLett.124.062001} {\bibfield  {journal} {\bibinfo  {journal} {Phys. Rev. Lett.}\ }\textbf {\bibinfo {volume} {124}},\ \bibinfo {pages} {062001} (\bibinfo {year} {2020})},\ \Eprint {http://arxiv.org/abs/1904.11974} {arXiv:1904.11974 [hep-ph]} \BibitemShut {NoStop}%
\bibitem [{\citenamefont {Hentschinski}\ and\ \citenamefont {Kutak}(2022)}]{Hentschinski:2021aux}%
  \BibitemOpen
  \bibfield  {author} {\bibinfo {author} {\bibfnamefont {M.}~\bibnamefont {Hentschinski}}\ and\ \bibinfo {author} {\bibfnamefont {K.}~\bibnamefont {Kutak}},\ }\href {\doibase 10.1140/epjc/s10052-022-10056-y} {\bibfield  {journal} {\bibinfo  {journal} {Eur. Phys. J. C}\ }\textbf {\bibinfo {volume} {82}},\ \bibinfo {pages} {111} (\bibinfo {year} {2022})},\ \Eprint {http://arxiv.org/abs/2110.06156} {arXiv:2110.06156 [hep-ph]} \BibitemShut {NoStop}%
\bibitem [{\citenamefont {Hentschinski}\ \emph {et~al.}(2022)\citenamefont {Hentschinski}, \citenamefont {Kutak},\ and\ \citenamefont {Straka}}]{Hentschinski:2022rsa}%
  \BibitemOpen
  \bibfield  {author} {\bibinfo {author} {\bibfnamefont {M.}~\bibnamefont {Hentschinski}}, \bibinfo {author} {\bibfnamefont {K.}~\bibnamefont {Kutak}}, \ and\ \bibinfo {author} {\bibfnamefont {R.}~\bibnamefont {Straka}},\ }\href {\doibase 10.1140/epjc/s10052-022-11122-1} {\bibfield  {journal} {\bibinfo  {journal} {Eur. Phys. J. C}\ }\textbf {\bibinfo {volume} {82}},\ \bibinfo {pages} {1147} (\bibinfo {year} {2022})},\ \Eprint {http://arxiv.org/abs/2207.09430} {arXiv:2207.09430 [hep-ph]} \BibitemShut {NoStop}%
\bibitem [{\citenamefont {Hentschinski}\ \emph {et~al.}(2023)\citenamefont {Hentschinski}, \citenamefont {Kharzeev}, \citenamefont {Kutak},\ and\ \citenamefont {Tu}}]{Hentschinski:2023izh}%
  \BibitemOpen
  \bibfield  {author} {\bibinfo {author} {\bibfnamefont {M.}~\bibnamefont {Hentschinski}}, \bibinfo {author} {\bibfnamefont {D.~E.}\ \bibnamefont {Kharzeev}}, \bibinfo {author} {\bibfnamefont {K.}~\bibnamefont {Kutak}}, \ and\ \bibinfo {author} {\bibfnamefont {Z.}~\bibnamefont {Tu}},\ }\href {\doibase 10.1103/PhysRevLett.131.241901} {\bibfield  {journal} {\bibinfo  {journal} {Phys. Rev. Lett.}\ }\textbf {\bibinfo {volume} {131}},\ \bibinfo {pages} {241901} (\bibinfo {year} {2023})},\ \Eprint {http://arxiv.org/abs/2305.03069} {arXiv:2305.03069 [hep-ph]} \BibitemShut {NoStop}%
\bibitem [{\citenamefont {Lipatov}(1986)}]{Lipatov:1985uk}%
  \BibitemOpen
  \bibfield  {author} {\bibinfo {author} {\bibfnamefont {L.~N.}\ \bibnamefont {Lipatov}},\ }\href@noop {} {\bibfield  {journal} {\bibinfo  {journal} {Sov. Phys. JETP}\ }\textbf {\bibinfo {volume} {63}},\ \bibinfo {pages} {904} (\bibinfo {year} {1986})}\BibitemShut {NoStop}%
\bibitem [{\citenamefont {Lipatov}(1990)}]{Lipatov:1990zb}%
  \BibitemOpen
  \bibfield  {author} {\bibinfo {author} {\bibfnamefont {L.~N.}\ \bibnamefont {Lipatov}},\ }\href {\doibase 10.1016/0370-2693(90)90937-2} {\bibfield  {journal} {\bibinfo  {journal} {Phys. Lett. B}\ }\textbf {\bibinfo {volume} {251}},\ \bibinfo {pages} {284} (\bibinfo {year} {1990})}\BibitemShut {NoStop}%
\bibitem [{\citenamefont {Lipatov}(1994)}]{Lipatov:1993yb}%
  \BibitemOpen
  \bibfield  {author} {\bibinfo {author} {\bibfnamefont {L.~N.}\ \bibnamefont {Lipatov}},\ }\href@noop {} {\bibfield  {journal} {\bibinfo  {journal} {JETP Lett.}\ }\textbf {\bibinfo {volume} {59}},\ \bibinfo {pages} {596} (\bibinfo {year} {1994})},\ \Eprint {http://arxiv.org/abs/hep-th/9311037} {arXiv:hep-th/9311037} \BibitemShut {NoStop}%
\bibitem [{\citenamefont {Braun}\ \emph {et~al.}(2003)\citenamefont {Braun}, \citenamefont {Korchemsky},\ and\ \citenamefont {M\"uller}}]{Braun:2003rp}%
  \BibitemOpen
  \bibfield  {author} {\bibinfo {author} {\bibfnamefont {V.~M.}\ \bibnamefont {Braun}}, \bibinfo {author} {\bibfnamefont {G.~P.}\ \bibnamefont {Korchemsky}}, \ and\ \bibinfo {author} {\bibfnamefont {D.}~\bibnamefont {M\"uller}},\ }\href {\doibase 10.1016/S0146-6410(03)90004-4} {\bibfield  {journal} {\bibinfo  {journal} {Prog. Part. Nucl. Phys.}\ }\textbf {\bibinfo {volume} {51}},\ \bibinfo {pages} {311} (\bibinfo {year} {2003})},\ \Eprint {http://arxiv.org/abs/hep-ph/0306057} {arXiv:hep-ph/0306057} \BibitemShut {NoStop}%
\bibitem [{\citenamefont {G\"ursoy}\ \emph {et~al.}(2023)\citenamefont {G\"ursoy}, \citenamefont {Kharzeev},\ and\ \citenamefont {Pedraza}}]{Gursoy:2023hge}%
  \BibitemOpen
  \bibfield  {author} {\bibinfo {author} {\bibfnamefont {U.}~\bibnamefont {G\"ursoy}}, \bibinfo {author} {\bibfnamefont {D.~E.}\ \bibnamefont {Kharzeev}}, \ and\ \bibinfo {author} {\bibfnamefont {J.~F.}\ \bibnamefont {Pedraza}},\ }\href@noop {} {\  (\bibinfo {year} {2023})},\ \Eprint {http://arxiv.org/abs/2306.16145} {arXiv:2306.16145 [hep-th]} \BibitemShut {NoStop}%
\bibitem [{\citenamefont {Caputa}\ and\ \citenamefont {Kutak}(2024)}]{Caputa:2024xkp}%
  \BibitemOpen
  \bibfield  {author} {\bibinfo {author} {\bibfnamefont {P.}~\bibnamefont {Caputa}}\ and\ \bibinfo {author} {\bibfnamefont {K.}~\bibnamefont {Kutak}},\ }\href@noop {} {\  (\bibinfo {year} {2024})},\ \Eprint {http://arxiv.org/abs/2404.07657} {arXiv:2404.07657 [hep-ph]} \BibitemShut {NoStop}%
\bibitem [{\citenamefont {Dumitru}\ \emph {et~al.}(2023)\citenamefont {Dumitru}, \citenamefont {Kovner},\ and\ \citenamefont {Skokov}}]{Dumitru:2023qee}%
  \BibitemOpen
  \bibfield  {author} {\bibinfo {author} {\bibfnamefont {A.}~\bibnamefont {Dumitru}}, \bibinfo {author} {\bibfnamefont {A.}~\bibnamefont {Kovner}}, \ and\ \bibinfo {author} {\bibfnamefont {V.~V.}\ \bibnamefont {Skokov}},\ }\href {\doibase 10.1103/PhysRevD.108.014014} {\bibfield  {journal} {\bibinfo  {journal} {Phys. Rev. D}\ }\textbf {\bibinfo {volume} {108}},\ \bibinfo {pages} {014014} (\bibinfo {year} {2023})},\ \Eprint {http://arxiv.org/abs/2304.08564} {arXiv:2304.08564 [hep-ph]} \BibitemShut {NoStop}%
\bibitem [{\citenamefont {Kutak}(2011)}]{Kutak:2011rb}%
  \BibitemOpen
  \bibfield  {author} {\bibinfo {author} {\bibfnamefont {K.}~\bibnamefont {Kutak}},\ }\href {\doibase 10.1016/j.physletb.2011.09.113} {\bibfield  {journal} {\bibinfo  {journal} {Phys. Lett. B}\ }\textbf {\bibinfo {volume} {705}},\ \bibinfo {pages} {217} (\bibinfo {year} {2011})},\ \Eprint {http://arxiv.org/abs/1103.3654} {arXiv:1103.3654 [hep-ph]} \BibitemShut {NoStop}%
\bibitem [{\citenamefont {Peschanski}(2013)}]{Peschanski:2012cw}%
  \BibitemOpen
  \bibfield  {author} {\bibinfo {author} {\bibfnamefont {R.}~\bibnamefont {Peschanski}},\ }\href {\doibase 10.1103/PhysRevD.87.034042} {\bibfield  {journal} {\bibinfo  {journal} {Phys. Rev. D}\ }\textbf {\bibinfo {volume} {87}},\ \bibinfo {pages} {034042} (\bibinfo {year} {2013})},\ \Eprint {http://arxiv.org/abs/1211.6911} {arXiv:1211.6911 [hep-ph]} \BibitemShut {NoStop}%
\bibitem [{\citenamefont {Armesto}\ \emph {et~al.}(2019)\citenamefont {Armesto}, \citenamefont {Dominguez}, \citenamefont {Kovner}, \citenamefont {Lublinsky},\ and\ \citenamefont {Skokov}}]{Armesto:2019mna}%
  \BibitemOpen
  \bibfield  {author} {\bibinfo {author} {\bibfnamefont {N.}~\bibnamefont {Armesto}}, \bibinfo {author} {\bibfnamefont {F.}~\bibnamefont {Dominguez}}, \bibinfo {author} {\bibfnamefont {A.}~\bibnamefont {Kovner}}, \bibinfo {author} {\bibfnamefont {M.}~\bibnamefont {Lublinsky}}, \ and\ \bibinfo {author} {\bibfnamefont {V.}~\bibnamefont {Skokov}},\ }\href {\doibase 10.1007/JHEP05(2019)025} {\bibfield  {journal} {\bibinfo  {journal} {JHEP}\ }\textbf {\bibinfo {volume} {05}},\ \bibinfo {pages} {025} (\bibinfo {year} {2019})},\ \Eprint {http://arxiv.org/abs/1901.08080} {arXiv:1901.08080 [hep-ph]} \BibitemShut {NoStop}%
\bibitem [{\citenamefont {Neill}\ and\ \citenamefont {Waalewijn}(2019)}]{Neill:2018uqw}%
  \BibitemOpen
  \bibfield  {author} {\bibinfo {author} {\bibfnamefont {D.}~\bibnamefont {Neill}}\ and\ \bibinfo {author} {\bibfnamefont {W.~J.}\ \bibnamefont {Waalewijn}},\ }\href {\doibase 10.1103/PhysRevLett.123.142001} {\bibfield  {journal} {\bibinfo  {journal} {Phys. Rev. Lett.}\ }\textbf {\bibinfo {volume} {123}},\ \bibinfo {pages} {142001} (\bibinfo {year} {2019})},\ \Eprint {http://arxiv.org/abs/1811.01021} {arXiv:1811.01021 [hep-ph]} \BibitemShut {NoStop}%
\bibitem [{\citenamefont {Kovner}\ \emph {et~al.}(2019)\citenamefont {Kovner}, \citenamefont {Lublinsky},\ and\ \citenamefont {Serino}}]{Kovner:2018rbf}%
  \BibitemOpen
  \bibfield  {author} {\bibinfo {author} {\bibfnamefont {A.}~\bibnamefont {Kovner}}, \bibinfo {author} {\bibfnamefont {M.}~\bibnamefont {Lublinsky}}, \ and\ \bibinfo {author} {\bibfnamefont {M.}~\bibnamefont {Serino}},\ }\href {\doibase 10.1016/j.physletb.2018.10.043} {\bibfield  {journal} {\bibinfo  {journal} {Phys. Lett. B}\ }\textbf {\bibinfo {volume} {792}},\ \bibinfo {pages} {4} (\bibinfo {year} {2019})},\ \Eprint {http://arxiv.org/abs/1806.01089} {arXiv:1806.01089 [hep-ph]} \BibitemShut {NoStop}%
\bibitem [{\citenamefont {Chachamis}\ \emph {et~al.}(2024)\citenamefont {Chachamis}, \citenamefont {Hentschinski},\ and\ \citenamefont {Sabio~Vera}}]{Chachamis:2023omp}%
  \BibitemOpen
  \bibfield  {author} {\bibinfo {author} {\bibfnamefont {G.}~\bibnamefont {Chachamis}}, \bibinfo {author} {\bibfnamefont {M.}~\bibnamefont {Hentschinski}}, \ and\ \bibinfo {author} {\bibfnamefont {A.}~\bibnamefont {Sabio~Vera}},\ }\href {\doibase 10.1103/PhysRevD.109.054015} {\bibfield  {journal} {\bibinfo  {journal} {Phys. Rev. D}\ }\textbf {\bibinfo {volume} {109}},\ \bibinfo {pages} {054015} (\bibinfo {year} {2024})},\ \Eprint {http://arxiv.org/abs/2312.16743} {arXiv:2312.16743 [hep-th]} \BibitemShut {NoStop}%
\bibitem [{\citenamefont {Liu}\ \emph {et~al.}(2022{\natexlab{a}})\citenamefont {Liu}, \citenamefont {Nowak},\ and\ \citenamefont {Zahed}}]{Liu:2022hto}%
  \BibitemOpen
  \bibfield  {author} {\bibinfo {author} {\bibfnamefont {Y.}~\bibnamefont {Liu}}, \bibinfo {author} {\bibfnamefont {M.~A.}\ \bibnamefont {Nowak}}, \ and\ \bibinfo {author} {\bibfnamefont {I.}~\bibnamefont {Zahed}},\ }\href {\doibase 10.1103/PhysRevD.105.114028} {\bibfield  {journal} {\bibinfo  {journal} {Phys. Rev. D}\ }\textbf {\bibinfo {volume} {105}},\ \bibinfo {pages} {114028} (\bibinfo {year} {2022}{\natexlab{a}})},\ \Eprint {http://arxiv.org/abs/2203.00739} {arXiv:2203.00739 [hep-ph]} \BibitemShut {NoStop}%
\bibitem [{\citenamefont {Liu}\ \emph {et~al.}(2023{\natexlab{a}})\citenamefont {Liu}, \citenamefont {Nowak},\ and\ \citenamefont {Zahed}}]{Liu:2023eve}%
  \BibitemOpen
  \bibfield  {author} {\bibinfo {author} {\bibfnamefont {Y.}~\bibnamefont {Liu}}, \bibinfo {author} {\bibfnamefont {M.~A.}\ \bibnamefont {Nowak}}, \ and\ \bibinfo {author} {\bibfnamefont {I.}~\bibnamefont {Zahed}},\ }\href@noop {} {\  (\bibinfo {year} {2023}{\natexlab{a}})},\ \Eprint {http://arxiv.org/abs/2302.01380} {arXiv:2302.01380 [hep-ph]} \BibitemShut {NoStop}%
\bibitem [{\citenamefont {Liu}\ \emph {et~al.}(2022{\natexlab{b}})\citenamefont {Liu}, \citenamefont {Nowak},\ and\ \citenamefont {Zahed}}]{Liu:2022bru}%
  \BibitemOpen
  \bibfield  {author} {\bibinfo {author} {\bibfnamefont {Y.}~\bibnamefont {Liu}}, \bibinfo {author} {\bibfnamefont {M.~A.}\ \bibnamefont {Nowak}}, \ and\ \bibinfo {author} {\bibfnamefont {I.}~\bibnamefont {Zahed}},\ }\href@noop {} {\  (\bibinfo {year} {2022}{\natexlab{b}})},\ \Eprint {http://arxiv.org/abs/2211.05169} {arXiv:2211.05169 [hep-ph]} \BibitemShut {NoStop}%
\bibitem [{\citenamefont {Liu}\ \emph {et~al.}(2023{\natexlab{b}})\citenamefont {Liu}, \citenamefont {Nowak},\ and\ \citenamefont {Zahed}}]{Liu:2022qqf}%
  \BibitemOpen
  \bibfield  {author} {\bibinfo {author} {\bibfnamefont {Y.}~\bibnamefont {Liu}}, \bibinfo {author} {\bibfnamefont {M.~A.}\ \bibnamefont {Nowak}}, \ and\ \bibinfo {author} {\bibfnamefont {I.}~\bibnamefont {Zahed}},\ }\href {\doibase 10.1103/PhysRevD.107.054010} {\bibfield  {journal} {\bibinfo  {journal} {Phys. Rev. D}\ }\textbf {\bibinfo {volume} {107}},\ \bibinfo {pages} {054010} (\bibinfo {year} {2023}{\natexlab{b}})},\ \Eprint {http://arxiv.org/abs/2205.06724} {arXiv:2205.06724 [hep-ph]} \BibitemShut {NoStop}%
\bibitem [{\citenamefont {Stoffers}\ and\ \citenamefont {Zahed}(2013)}]{Stoffers:2012mn}%
  \BibitemOpen
  \bibfield  {author} {\bibinfo {author} {\bibfnamefont {A.}~\bibnamefont {Stoffers}}\ and\ \bibinfo {author} {\bibfnamefont {I.}~\bibnamefont {Zahed}},\ }\href {\doibase 10.1103/PhysRevD.88.025038} {\bibfield  {journal} {\bibinfo  {journal} {Phys. Rev. D}\ }\textbf {\bibinfo {volume} {88}},\ \bibinfo {pages} {025038} (\bibinfo {year} {2013})},\ \Eprint {http://arxiv.org/abs/1211.3077} {arXiv:1211.3077 [nucl-th]} \BibitemShut {NoStop}%
\bibitem [{\citenamefont {Asadi}\ and\ \citenamefont {Vaidya}(2023)}]{Asadi:2023bat}%
  \BibitemOpen
  \bibfield  {author} {\bibinfo {author} {\bibfnamefont {P.}~\bibnamefont {Asadi}}\ and\ \bibinfo {author} {\bibfnamefont {V.}~\bibnamefont {Vaidya}},\ }\href {\doibase 10.1103/PhysRevD.108.014036} {\bibfield  {journal} {\bibinfo  {journal} {Phys. Rev. D}\ }\textbf {\bibinfo {volume} {108}},\ \bibinfo {pages} {014036} (\bibinfo {year} {2023})},\ \Eprint {http://arxiv.org/abs/2301.03611} {arXiv:2301.03611 [hep-th]} \BibitemShut {NoStop}%
\bibitem [{\citenamefont {Kou}\ \emph {et~al.}(2022)\citenamefont {Kou}, \citenamefont {Wang},\ and\ \citenamefont {Chen}}]{Kou:2022dkw}%
  \BibitemOpen
  \bibfield  {author} {\bibinfo {author} {\bibfnamefont {W.}~\bibnamefont {Kou}}, \bibinfo {author} {\bibfnamefont {X.}~\bibnamefont {Wang}}, \ and\ \bibinfo {author} {\bibfnamefont {X.}~\bibnamefont {Chen}},\ }\href {\doibase 10.1103/PhysRevD.106.096027} {\bibfield  {journal} {\bibinfo  {journal} {Phys. Rev. D}\ }\textbf {\bibinfo {volume} {106}},\ \bibinfo {pages} {096027} (\bibinfo {year} {2022})},\ \Eprint {http://arxiv.org/abs/2208.07521} {arXiv:2208.07521 [hep-ph]} \BibitemShut {NoStop}%
\bibitem [{\citenamefont {Kutak}(2023)}]{Kutak:2023cwg}%
  \BibitemOpen
  \bibfield  {author} {\bibinfo {author} {\bibfnamefont {K.}~\bibnamefont {Kutak}},\ }\href@noop {} {\  (\bibinfo {year} {2023})},\ \Eprint {http://arxiv.org/abs/2310.18510} {arXiv:2310.18510 [hep-ph]} \BibitemShut {NoStop}%
\bibitem [{\citenamefont {Ehlers}(2022)}]{Ehlers:2022oke}%
  \BibitemOpen
  \bibfield  {author} {\bibinfo {author} {\bibfnamefont {P.~J.}\ \bibnamefont {Ehlers}},\ }\emph {\bibinfo {title} {{Entanglement between Quarks in Hadrons}}},\ \href@noop {} {Ph.D. thesis},\ \bibinfo  {school} {Washington U., Seattle} (\bibinfo {year} {2022})\BibitemShut {NoStop}%
\bibitem [{\citenamefont {Ehlers}(2023)}]{Ehlers:2022oal}%
  \BibitemOpen
  \bibfield  {author} {\bibinfo {author} {\bibfnamefont {P.~J.}\ \bibnamefont {Ehlers}},\ }\href {\doibase 10.1016/j.aop.2023.169290} {\bibfield  {journal} {\bibinfo  {journal} {Annals Phys.}\ }\textbf {\bibinfo {volume} {452}},\ \bibinfo {pages} {169290} (\bibinfo {year} {2023})},\ \Eprint {http://arxiv.org/abs/2209.09867} {arXiv:2209.09867 [hep-ph]} \BibitemShut {NoStop}%
\bibitem [{\citenamefont {Florio}\ \emph {et~al.}(2024)\citenamefont {Florio}, \citenamefont {Frenklakh}, \citenamefont {Ikeda}, \citenamefont {Kharzeev}, \citenamefont {Korepin}, \citenamefont {Shi},\ and\ \citenamefont {Yu}}]{Florio:2024aix}%
  \BibitemOpen
  \bibfield  {author} {\bibinfo {author} {\bibfnamefont {A.}~\bibnamefont {Florio}}, \bibinfo {author} {\bibfnamefont {D.}~\bibnamefont {Frenklakh}}, \bibinfo {author} {\bibfnamefont {K.}~\bibnamefont {Ikeda}}, \bibinfo {author} {\bibfnamefont {D.~E.}\ \bibnamefont {Kharzeev}}, \bibinfo {author} {\bibfnamefont {V.}~\bibnamefont {Korepin}}, \bibinfo {author} {\bibfnamefont {S.}~\bibnamefont {Shi}}, \ and\ \bibinfo {author} {\bibfnamefont {K.}~\bibnamefont {Yu}},\ }\href@noop {} {\  (\bibinfo {year} {2024})},\ \Eprint {http://arxiv.org/abs/2404.00087} {arXiv:2404.00087 [hep-ph]} \BibitemShut {NoStop}%
\bibitem [{\citenamefont {Grieninger}\ \emph {et~al.}(2024)\citenamefont {Grieninger}, \citenamefont {Ikeda},\ and\ \citenamefont {Kharzeev}}]{Grieninger:2023knz}%
  \BibitemOpen
  \bibfield  {author} {\bibinfo {author} {\bibfnamefont {S.}~\bibnamefont {Grieninger}}, \bibinfo {author} {\bibfnamefont {K.}~\bibnamefont {Ikeda}}, \ and\ \bibinfo {author} {\bibfnamefont {D.~E.}\ \bibnamefont {Kharzeev}},\ }\href {\doibase 10.1007/JHEP05(2024)030} {\bibfield  {journal} {\bibinfo  {journal} {JHEP}\ }\textbf {\bibinfo {volume} {05}},\ \bibinfo {pages} {030} (\bibinfo {year} {2024})},\ \Eprint {http://arxiv.org/abs/2312.08534} {arXiv:2312.08534 [hep-th]} \BibitemShut {NoStop}%
\bibitem [{\citenamefont {Grieninger}\ \emph {et~al.}(2023{\natexlab{b}})\citenamefont {Grieninger}, \citenamefont {Kharzeev},\ and\ \citenamefont {Zahed}}]{Grieninger:2023pyb}%
  \BibitemOpen
  \bibfield  {author} {\bibinfo {author} {\bibfnamefont {S.}~\bibnamefont {Grieninger}}, \bibinfo {author} {\bibfnamefont {D.~E.}\ \bibnamefont {Kharzeev}}, \ and\ \bibinfo {author} {\bibfnamefont {I.}~\bibnamefont {Zahed}},\ }\href {\doibase 10.1103/PhysRevD.108.126014} {\bibfield  {journal} {\bibinfo  {journal} {Phys. Rev. D}\ }\textbf {\bibinfo {volume} {108}},\ \bibinfo {pages} {126014} (\bibinfo {year} {2023}{\natexlab{b}})},\ \Eprint {http://arxiv.org/abs/2310.12042} {arXiv:2310.12042 [hep-th]} \BibitemShut {NoStop}%
\bibitem [{\citenamefont {Ikeda}\ \emph {et~al.}(2023)\citenamefont {Ikeda}, \citenamefont {Kharzeev}, \citenamefont {Meyer},\ and\ \citenamefont {Shi}}]{Ikeda:2023zil}%
  \BibitemOpen
  \bibfield  {author} {\bibinfo {author} {\bibfnamefont {K.}~\bibnamefont {Ikeda}}, \bibinfo {author} {\bibfnamefont {D.~E.}\ \bibnamefont {Kharzeev}}, \bibinfo {author} {\bibfnamefont {R.}~\bibnamefont {Meyer}}, \ and\ \bibinfo {author} {\bibfnamefont {S.}~\bibnamefont {Shi}},\ }\href {\doibase 10.1103/PhysRevD.108.L091501} {\bibfield  {journal} {\bibinfo  {journal} {Phys. Rev. D}\ }\textbf {\bibinfo {volume} {108}},\ \bibinfo {pages} {L091501} (\bibinfo {year} {2023})},\ \Eprint {http://arxiv.org/abs/2305.00996} {arXiv:2305.00996 [hep-ph]} \BibitemShut {NoStop}%
\bibitem [{\citenamefont {Brandenburg}\ \emph {et~al.}(2024)\citenamefont {Brandenburg}, \citenamefont {Duan}, \citenamefont {Tu}, \citenamefont {Venugopalan},\ and\ \citenamefont {Xu}}]{Brandenburg:2024ksp}%
  \BibitemOpen
  \bibfield  {author} {\bibinfo {author} {\bibfnamefont {J.~D.}\ \bibnamefont {Brandenburg}}, \bibinfo {author} {\bibfnamefont {H.}~\bibnamefont {Duan}}, \bibinfo {author} {\bibfnamefont {Z.}~\bibnamefont {Tu}}, \bibinfo {author} {\bibfnamefont {R.}~\bibnamefont {Venugopalan}}, \ and\ \bibinfo {author} {\bibfnamefont {Z.}~\bibnamefont {Xu}},\ }\href@noop {} {\  (\bibinfo {year} {2024})},\ \Eprint {http://arxiv.org/abs/2407.15945} {arXiv:2407.15945 [hep-ph]} \BibitemShut {NoStop}%
\bibitem [{\citenamefont {Berges}\ \emph {et~al.}(2018{\natexlab{a}})\citenamefont {Berges}, \citenamefont {Floerchinger},\ and\ \citenamefont {Venugopalan}}]{Berges:2017hne}%
  \BibitemOpen
  \bibfield  {author} {\bibinfo {author} {\bibfnamefont {J.}~\bibnamefont {Berges}}, \bibinfo {author} {\bibfnamefont {S.}~\bibnamefont {Floerchinger}}, \ and\ \bibinfo {author} {\bibfnamefont {R.}~\bibnamefont {Venugopalan}},\ }\href {\doibase 10.1007/JHEP04(2018)145} {\bibfield  {journal} {\bibinfo  {journal} {JHEP}\ }\textbf {\bibinfo {volume} {04}},\ \bibinfo {pages} {145} (\bibinfo {year} {2018}{\natexlab{a}})},\ \Eprint {http://arxiv.org/abs/1712.09362} {arXiv:1712.09362 [hep-th]} \BibitemShut {NoStop}%
\bibitem [{\citenamefont {Berges}\ \emph {et~al.}(2018{\natexlab{b}})\citenamefont {Berges}, \citenamefont {Floerchinger},\ and\ \citenamefont {Venugopalan}}]{Berges:2017zws}%
  \BibitemOpen
  \bibfield  {author} {\bibinfo {author} {\bibfnamefont {J.}~\bibnamefont {Berges}}, \bibinfo {author} {\bibfnamefont {S.}~\bibnamefont {Floerchinger}}, \ and\ \bibinfo {author} {\bibfnamefont {R.}~\bibnamefont {Venugopalan}},\ }\href {\doibase 10.1016/j.physletb.2018.01.068} {\bibfield  {journal} {\bibinfo  {journal} {Phys. Lett. B}\ }\textbf {\bibinfo {volume} {778}},\ \bibinfo {pages} {442} (\bibinfo {year} {2018}{\natexlab{b}})},\ \Eprint {http://arxiv.org/abs/1707.05338} {arXiv:1707.05338 [hep-ph]} \BibitemShut {NoStop}%
\bibitem [{\citenamefont {Dumitru}\ and\ \citenamefont {Kolbusz}(2023)}]{Dumitru:2023fih}%
  \BibitemOpen
  \bibfield  {author} {\bibinfo {author} {\bibfnamefont {A.}~\bibnamefont {Dumitru}}\ and\ \bibinfo {author} {\bibfnamefont {E.}~\bibnamefont {Kolbusz}},\ }\href {\doibase 10.1103/PhysRevD.108.034011} {\bibfield  {journal} {\bibinfo  {journal} {Phys. Rev. D}\ }\textbf {\bibinfo {volume} {108}},\ \bibinfo {pages} {034011} (\bibinfo {year} {2023})},\ \Eprint {http://arxiv.org/abs/2303.07408} {arXiv:2303.07408 [hep-ph]} \BibitemShut {NoStop}%
\bibitem [{\citenamefont {Dumitru}\ and\ \citenamefont {Kolbusz}(2022)}]{Dumitru:2022tud}%
  \BibitemOpen
  \bibfield  {author} {\bibinfo {author} {\bibfnamefont {A.}~\bibnamefont {Dumitru}}\ and\ \bibinfo {author} {\bibfnamefont {E.}~\bibnamefont {Kolbusz}},\ }\href {\doibase 10.1103/PhysRevD.105.074030} {\bibfield  {journal} {\bibinfo  {journal} {Phys. Rev. D}\ }\textbf {\bibinfo {volume} {105}},\ \bibinfo {pages} {074030} (\bibinfo {year} {2022})},\ \Eprint {http://arxiv.org/abs/2202.01803} {arXiv:2202.01803 [hep-ph]} \BibitemShut {NoStop}%
\bibitem [{\citenamefont {Ramos}\ and\ \citenamefont {Machado}(2022)}]{Ramos:2022gia}%
  \BibitemOpen
  \bibfield  {author} {\bibinfo {author} {\bibfnamefont {G.~S.}\ \bibnamefont {Ramos}}\ and\ \bibinfo {author} {\bibfnamefont {M.~V.~T.}\ \bibnamefont {Machado}},\ }\href {\doibase 10.1103/PhysRevD.105.094009} {\bibfield  {journal} {\bibinfo  {journal} {Phys. Rev. D}\ }\textbf {\bibinfo {volume} {105}},\ \bibinfo {pages} {094009} (\bibinfo {year} {2022})},\ \Eprint {http://arxiv.org/abs/2203.10986} {arXiv:2203.10986 [hep-ph]} \BibitemShut {NoStop}%
\bibitem [{\citenamefont {Moriggi}\ \emph {et~al.}(2024)\citenamefont {Moriggi}, \citenamefont {Ramos},\ and\ \citenamefont {Machado}}]{Moriggi:2024tbr}%
  \BibitemOpen
  \bibfield  {author} {\bibinfo {author} {\bibfnamefont {L.~S.}\ \bibnamefont {Moriggi}}, \bibinfo {author} {\bibfnamefont {G.~S.}\ \bibnamefont {Ramos}}, \ and\ \bibinfo {author} {\bibfnamefont {M.~V.~T.}\ \bibnamefont {Machado}},\ }\href@noop {} {\  (\bibinfo {year} {2024})},\ \Eprint {http://arxiv.org/abs/2405.01712} {arXiv:2405.01712 [hep-ph]} \BibitemShut {NoStop}%
\bibitem [{\citenamefont {Ramos}\ and\ \citenamefont {Machado}(2020)}]{Ramos:2020kaj}%
  \BibitemOpen
  \bibfield  {author} {\bibinfo {author} {\bibfnamefont {G.~S.}\ \bibnamefont {Ramos}}\ and\ \bibinfo {author} {\bibfnamefont {M.~V.~T.}\ \bibnamefont {Machado}},\ }\href {\doibase 10.1103/PhysRevD.101.074040} {\bibfield  {journal} {\bibinfo  {journal} {Phys. Rev. D}\ }\textbf {\bibinfo {volume} {101}},\ \bibinfo {pages} {074040} (\bibinfo {year} {2020})},\ \Eprint {http://arxiv.org/abs/2003.05008} {arXiv:2003.05008 [hep-ph]} \BibitemShut {NoStop}%
\bibitem [{\citenamefont {Dosch}\ \emph {et~al.}(2024)\citenamefont {Dosch}, \citenamefont {de~Teramond},\ and\ \citenamefont {Brodsky}}]{Dosch:2023bxj}%
  \BibitemOpen
  \bibfield  {author} {\bibinfo {author} {\bibfnamefont {H.~G.}\ \bibnamefont {Dosch}}, \bibinfo {author} {\bibfnamefont {G.~F.}\ \bibnamefont {de~Teramond}}, \ and\ \bibinfo {author} {\bibfnamefont {S.~J.}\ \bibnamefont {Brodsky}},\ }\href {\doibase 10.1016/j.physletb.2024.138521} {\bibfield  {journal} {\bibinfo  {journal} {Phys. Lett. B}\ }\textbf {\bibinfo {volume} {850}},\ \bibinfo {pages} {138521} (\bibinfo {year} {2024})},\ \Eprint {http://arxiv.org/abs/2304.14207} {arXiv:2304.14207 [hep-ph]} \BibitemShut {NoStop}%
\bibitem [{\citenamefont {Aid}\ \emph {et~al.}(1996)\citenamefont {Aid} \emph {et~al.}}]{H1:1996ovs}%
  \BibitemOpen
  \bibfield  {author} {\bibinfo {author} {\bibfnamefont {S.}~\bibnamefont {Aid}} \emph {et~al.} (\bibinfo {collaboration} {H1}),\ }\href {\doibase 10.1007/s002880050280} {\bibfield  {journal} {\bibinfo  {journal} {Z. Phys. C}\ }\textbf {\bibinfo {volume} {72}},\ \bibinfo {pages} {573} (\bibinfo {year} {1996})},\ \Eprint {http://arxiv.org/abs/hep-ex/9608011} {arXiv:hep-ex/9608011} \BibitemShut {NoStop}%
\bibitem [{\citenamefont {Andreev}\ \emph {et~al.}(2021)\citenamefont {Andreev} \emph {et~al.}}]{H1:2020zpd}%
  \BibitemOpen
  \bibfield  {author} {\bibinfo {author} {\bibfnamefont {V.}~\bibnamefont {Andreev}} \emph {et~al.} (\bibinfo {collaboration} {H1}),\ }\href {\doibase 10.1140/epjc/s10052-021-08896-1} {\bibfield  {journal} {\bibinfo  {journal} {Eur. Phys. J. C}\ }\textbf {\bibinfo {volume} {81}},\ \bibinfo {pages} {212} (\bibinfo {year} {2021})},\ \Eprint {http://arxiv.org/abs/2011.01812} {arXiv:2011.01812 [hep-ex]} \BibitemShut {NoStop}%
\bibitem [{\citenamefont {Mueller}\ and\ \citenamefont {Qiu}(1986)}]{Mueller:1985wy}%
  \BibitemOpen
  \bibfield  {author} {\bibinfo {author} {\bibfnamefont {A.~H.}\ \bibnamefont {Mueller}}\ and\ \bibinfo {author} {\bibfnamefont {J.-w.}\ \bibnamefont {Qiu}},\ }\href {\doibase 10.1016/0550-3213(86)90164-1} {\bibfield  {journal} {\bibinfo  {journal} {Nucl. Phys. B}\ }\textbf {\bibinfo {volume} {268}},\ \bibinfo {pages} {427} (\bibinfo {year} {1986})}\BibitemShut {NoStop}%
\bibitem [{\citenamefont {Kovchegov}\ and\ \citenamefont {Levin}(2013)}]{Kovchegov:2012mbw}%
  \BibitemOpen
  \bibfield  {author} {\bibinfo {author} {\bibfnamefont {Y.~V.}\ \bibnamefont {Kovchegov}}\ and\ \bibinfo {author} {\bibfnamefont {E.}~\bibnamefont {Levin}},\ }\href {\doibase 10.1017/9781009291446} {\emph {\bibinfo {title} {{Quantum Chromodynamics at High Energy}}}},\ Vol.~\bibinfo {volume} {33}\ (\bibinfo  {publisher} {Oxford University Press},\ \bibinfo {year} {2013})\BibitemShut {NoStop}%
\bibitem [{\citenamefont {Balitsky}(1996)}]{Balitsky:1995ub}%
  \BibitemOpen
  \bibfield  {author} {\bibinfo {author} {\bibfnamefont {I.}~\bibnamefont {Balitsky}},\ }\href {\doibase 10.1016/0550-3213(95)00638-9} {\bibfield  {journal} {\bibinfo  {journal} {Nucl. Phys. B}\ }\textbf {\bibinfo {volume} {463}},\ \bibinfo {pages} {99} (\bibinfo {year} {1996})},\ \Eprint {http://arxiv.org/abs/hep-ph/9509348} {arXiv:hep-ph/9509348} \BibitemShut {NoStop}%
\bibitem [{\citenamefont {Kovchegov}\ and\ \citenamefont {Levin}(2000)}]{Kovchegov:1999ji}%
  \BibitemOpen
  \bibfield  {author} {\bibinfo {author} {\bibfnamefont {Y.~V.}\ \bibnamefont {Kovchegov}}\ and\ \bibinfo {author} {\bibfnamefont {E.}~\bibnamefont {Levin}},\ }\href {\doibase 10.1016/S0550-3213(00)00125-5} {\bibfield  {journal} {\bibinfo  {journal} {Nucl. Phys. B}\ }\textbf {\bibinfo {volume} {577}},\ \bibinfo {pages} {221} (\bibinfo {year} {2000})},\ \Eprint {http://arxiv.org/abs/hep-ph/9911523} {arXiv:hep-ph/9911523} \BibitemShut {NoStop}%
\bibitem [{\citenamefont {Mueller}(1995)}]{Mueller:1994gb}%
  \BibitemOpen
  \bibfield  {author} {\bibinfo {author} {\bibfnamefont {A.~H.}\ \bibnamefont {Mueller}},\ }\href {\doibase 10.1016/0550-3213(94)00480-3} {\bibfield  {journal} {\bibinfo  {journal} {Nucl. Phys. B}\ }\textbf {\bibinfo {volume} {437}},\ \bibinfo {pages} {107} (\bibinfo {year} {1995})},\ \Eprint {http://arxiv.org/abs/hep-ph/9408245} {arXiv:hep-ph/9408245} \BibitemShut {NoStop}%
\bibitem [{\citenamefont {Levin}\ and\ \citenamefont {Lublinsky}(2004)}]{Levin:2003nc}%
  \BibitemOpen
  \bibfield  {author} {\bibinfo {author} {\bibfnamefont {E.}~\bibnamefont {Levin}}\ and\ \bibinfo {author} {\bibfnamefont {M.}~\bibnamefont {Lublinsky}},\ }\href {\doibase 10.1016/j.nuclphysa.2003.10.020} {\bibfield  {journal} {\bibinfo  {journal} {Nucl. Phys. A}\ }\textbf {\bibinfo {volume} {730}},\ \bibinfo {pages} {191} (\bibinfo {year} {2004})},\ \Eprint {http://arxiv.org/abs/hep-ph/0308279} {arXiv:hep-ph/0308279} \BibitemShut {NoStop}%
\bibitem [{\citenamefont {Kuraev}\ \emph {et~al.}(1976)\citenamefont {Kuraev}, \citenamefont {Lipatov},\ and\ \citenamefont {Fadin}}]{Kuraev:1976ge}%
  \BibitemOpen
  \bibfield  {author} {\bibinfo {author} {\bibfnamefont {E.~A.}\ \bibnamefont {Kuraev}}, \bibinfo {author} {\bibfnamefont {L.~N.}\ \bibnamefont {Lipatov}}, \ and\ \bibinfo {author} {\bibfnamefont {V.~S.}\ \bibnamefont {Fadin}},\ }\href@noop {} {\bibfield  {journal} {\bibinfo  {journal} {Sov. Phys. JETP}\ }\textbf {\bibinfo {volume} {44}},\ \bibinfo {pages} {443} (\bibinfo {year} {1976})}\BibitemShut {NoStop}%
\bibitem [{\citenamefont {Kuraev}\ \emph {et~al.}(1977)\citenamefont {Kuraev}, \citenamefont {Lipatov},\ and\ \citenamefont {Fadin}}]{Kuraev:1977fs}%
  \BibitemOpen
  \bibfield  {author} {\bibinfo {author} {\bibfnamefont {E.~A.}\ \bibnamefont {Kuraev}}, \bibinfo {author} {\bibfnamefont {L.~N.}\ \bibnamefont {Lipatov}}, \ and\ \bibinfo {author} {\bibfnamefont {V.~S.}\ \bibnamefont {Fadin}},\ }\href@noop {} {\bibfield  {journal} {\bibinfo  {journal} {Sov. Phys. JETP}\ }\textbf {\bibinfo {volume} {45}},\ \bibinfo {pages} {199} (\bibinfo {year} {1977})}\BibitemShut {NoStop}%
\bibitem [{\citenamefont {Balitsky}\ and\ \citenamefont {Lipatov}(1978)}]{Balitsky:1978ic}%
  \BibitemOpen
  \bibfield  {author} {\bibinfo {author} {\bibfnamefont {I.~I.}\ \bibnamefont {Balitsky}}\ and\ \bibinfo {author} {\bibfnamefont {L.~N.}\ \bibnamefont {Lipatov}},\ }\href@noop {} {\bibfield  {journal} {\bibinfo  {journal} {Sov. J. Nucl. Phys.}\ }\textbf {\bibinfo {volume} {28}},\ \bibinfo {pages} {822} (\bibinfo {year} {1978})}\BibitemShut {NoStop}%
\bibitem [{\citenamefont {Gribov}\ and\ \citenamefont {Lipatov}(1972)}]{Gribov:1972ri}%
  \BibitemOpen
  \bibfield  {author} {\bibinfo {author} {\bibfnamefont {V.~N.}\ \bibnamefont {Gribov}}\ and\ \bibinfo {author} {\bibfnamefont {L.~N.}\ \bibnamefont {Lipatov}},\ }\href@noop {} {\bibfield  {journal} {\bibinfo  {journal} {Sov. J. Nucl. Phys.}\ }\textbf {\bibinfo {volume} {15}},\ \bibinfo {pages} {438} (\bibinfo {year} {1972})}\BibitemShut {NoStop}%
\bibitem [{\citenamefont {Altarelli}\ and\ \citenamefont {Parisi}(1977)}]{Altarelli:1977zs}%
  \BibitemOpen
  \bibfield  {author} {\bibinfo {author} {\bibfnamefont {G.}~\bibnamefont {Altarelli}}\ and\ \bibinfo {author} {\bibfnamefont {G.}~\bibnamefont {Parisi}},\ }\href {\doibase 10.1016/0550-3213(77)90384-4} {\bibfield  {journal} {\bibinfo  {journal} {Nucl. Phys. B}\ }\textbf {\bibinfo {volume} {126}},\ \bibinfo {pages} {298} (\bibinfo {year} {1977})}\BibitemShut {NoStop}%
\bibitem [{\citenamefont {Dokshitzer}(1977)}]{Dokshitzer:1977sg}%
  \BibitemOpen
  \bibfield  {author} {\bibinfo {author} {\bibfnamefont {Y.~L.}\ \bibnamefont {Dokshitzer}},\ }\href@noop {} {\bibfield  {journal} {\bibinfo  {journal} {Sov. Phys. JETP}\ }\textbf {\bibinfo {volume} {46}},\ \bibinfo {pages} {641} (\bibinfo {year} {1977})}\BibitemShut {NoStop}%
\bibitem [{\citenamefont {Kharzeev}\ and\ \citenamefont {Levin}(2021)}]{Kharzeev:2021yyf}%
  \BibitemOpen
  \bibfield  {author} {\bibinfo {author} {\bibfnamefont {D.~E.}\ \bibnamefont {Kharzeev}}\ and\ \bibinfo {author} {\bibfnamefont {E.}~\bibnamefont {Levin}},\ }\href {\doibase 10.1103/PhysRevD.104.L031503} {\bibfield  {journal} {\bibinfo  {journal} {Phys. Rev. D}\ }\textbf {\bibinfo {volume} {104}},\ \bibinfo {pages} {L031503} (\bibinfo {year} {2021})},\ \Eprint {http://arxiv.org/abs/2102.09773} {arXiv:2102.09773 [hep-ph]} \BibitemShut {NoStop}%
\bibitem [{\citenamefont {Abramowicz}\ \emph {et~al.}(2015)\citenamefont {Abramowicz} \emph {et~al.}}]{H1:2015ubc}%
  \BibitemOpen
  \bibfield  {author} {\bibinfo {author} {\bibfnamefont {H.}~\bibnamefont {Abramowicz}} \emph {et~al.} (\bibinfo {collaboration} {H1, ZEUS}),\ }\href {\doibase 10.1140/epjc/s10052-015-3710-4} {\bibfield  {journal} {\bibinfo  {journal} {Eur. Phys. J. C}\ }\textbf {\bibinfo {volume} {75}},\ \bibinfo {pages} {580} (\bibinfo {year} {2015})},\ \Eprint {http://arxiv.org/abs/1506.06042} {arXiv:1506.06042 [hep-ex]} \BibitemShut {NoStop}%
\bibitem [{\citenamefont {Hentschinski}\ \emph {et~al.}(2013{\natexlab{a}})\citenamefont {Hentschinski}, \citenamefont {Sabio~Vera},\ and\ \citenamefont {Salas}}]{Hentschinski:2012kr}%
  \BibitemOpen
  \bibfield  {author} {\bibinfo {author} {\bibfnamefont {M.}~\bibnamefont {Hentschinski}}, \bibinfo {author} {\bibfnamefont {A.}~\bibnamefont {Sabio~Vera}}, \ and\ \bibinfo {author} {\bibfnamefont {C.}~\bibnamefont {Salas}},\ }\href {\doibase 10.1103/PhysRevLett.110.041601} {\bibfield  {journal} {\bibinfo  {journal} {Phys. Rev. Lett.}\ }\textbf {\bibinfo {volume} {110}},\ \bibinfo {pages} {041601} (\bibinfo {year} {2013}{\natexlab{a}})},\ \Eprint {http://arxiv.org/abs/1209.1353} {arXiv:1209.1353 [hep-ph]} \BibitemShut {NoStop}%
\bibitem [{\citenamefont {Hentschinski}\ \emph {et~al.}(2013{\natexlab{b}})\citenamefont {Hentschinski}, \citenamefont {Sabio~Vera},\ and\ \citenamefont {Salas}}]{Hentschinski:2013id}%
  \BibitemOpen
  \bibfield  {author} {\bibinfo {author} {\bibfnamefont {M.}~\bibnamefont {Hentschinski}}, \bibinfo {author} {\bibfnamefont {A.}~\bibnamefont {Sabio~Vera}}, \ and\ \bibinfo {author} {\bibfnamefont {C.}~\bibnamefont {Salas}},\ }\href {\doibase 10.1103/PhysRevD.87.076005} {\bibfield  {journal} {\bibinfo  {journal} {Phys. Rev. D}\ }\textbf {\bibinfo {volume} {87}},\ \bibinfo {pages} {076005} (\bibinfo {year} {2013}{\natexlab{b}})},\ \Eprint {http://arxiv.org/abs/1301.5283} {arXiv:1301.5283 [hep-ph]} \BibitemShut {NoStop}%
\bibitem [{\citenamefont {Chachamis}\ \emph {et~al.}(2015)\citenamefont {Chachamis}, \citenamefont {De\'ak}, \citenamefont {Hentschinski}, \citenamefont {Rodrigo},\ and\ \citenamefont {Sabio~Vera}}]{Chachamis:2015ona}%
  \BibitemOpen
  \bibfield  {author} {\bibinfo {author} {\bibfnamefont {G.}~\bibnamefont {Chachamis}}, \bibinfo {author} {\bibfnamefont {M.}~\bibnamefont {De\'ak}}, \bibinfo {author} {\bibfnamefont {M.}~\bibnamefont {Hentschinski}}, \bibinfo {author} {\bibfnamefont {G.}~\bibnamefont {Rodrigo}}, \ and\ \bibinfo {author} {\bibfnamefont {A.}~\bibnamefont {Sabio~Vera}},\ }\href {\doibase 10.1007/JHEP09(2015)123} {\bibfield  {journal} {\bibinfo  {journal} {JHEP}\ }\textbf {\bibinfo {volume} {09}},\ \bibinfo {pages} {123} (\bibinfo {year} {2015})},\ \Eprint {http://arxiv.org/abs/1507.05778} {arXiv:1507.05778 [hep-ph]} \BibitemShut {NoStop}%
\bibitem [{\citenamefont {Bautista}\ \emph {et~al.}(2016)\citenamefont {Bautista}, \citenamefont {Fernandez~Tellez},\ and\ \citenamefont {Hentschinski}}]{Bautista:2016xnp}%
  \BibitemOpen
  \bibfield  {author} {\bibinfo {author} {\bibfnamefont {I.}~\bibnamefont {Bautista}}, \bibinfo {author} {\bibfnamefont {A.}~\bibnamefont {Fernandez~Tellez}}, \ and\ \bibinfo {author} {\bibfnamefont {M.}~\bibnamefont {Hentschinski}},\ }\href {\doibase 10.1103/PhysRevD.94.054002} {\bibfield  {journal} {\bibinfo  {journal} {Phys. Rev. D}\ }\textbf {\bibinfo {volume} {94}},\ \bibinfo {pages} {054002} (\bibinfo {year} {2016})},\ \Eprint {http://arxiv.org/abs/1607.05203} {arXiv:1607.05203 [hep-ph]} \BibitemShut {NoStop}%
\bibitem [{\citenamefont {Fadin}\ and\ \citenamefont {Lipatov}(1998)}]{Fadin:1998py}%
  \BibitemOpen
  \bibfield  {author} {\bibinfo {author} {\bibfnamefont {V.~S.}\ \bibnamefont {Fadin}}\ and\ \bibinfo {author} {\bibfnamefont {L.~N.}\ \bibnamefont {Lipatov}},\ }\href {\doibase 10.1016/S0370-2693(98)00473-0} {\bibfield  {journal} {\bibinfo  {journal} {Phys. Lett. B}\ }\textbf {\bibinfo {volume} {429}},\ \bibinfo {pages} {127} (\bibinfo {year} {1998})},\ \Eprint {http://arxiv.org/abs/hep-ph/9802290} {arXiv:hep-ph/9802290} \BibitemShut {NoStop}%
\bibitem [{\citenamefont {Clark}\ \emph {et~al.}(2017)\citenamefont {Clark}, \citenamefont {Godat},\ and\ \citenamefont {Olness}}]{Clark:2016jgm}%
  \BibitemOpen
  \bibfield  {author} {\bibinfo {author} {\bibfnamefont {D.~B.}\ \bibnamefont {Clark}}, \bibinfo {author} {\bibfnamefont {E.}~\bibnamefont {Godat}}, \ and\ \bibinfo {author} {\bibfnamefont {F.~I.}\ \bibnamefont {Olness}},\ }\href {\doibase 10.1016/j.cpc.2017.03.004} {\bibfield  {journal} {\bibinfo  {journal} {Comput. Phys. Commun.}\ }\textbf {\bibinfo {volume} {216}},\ \bibinfo {pages} {126} (\bibinfo {year} {2017})},\ \Eprint {http://arxiv.org/abs/1605.08012} {arXiv:1605.08012 [hep-ph]} \BibitemShut {NoStop}%
\end{thebibliography}%

\end{document}